\documentclass{aa}

\renewcommand{\linenumbers}{}  
\nolinenumbers                 

\usepackage{graphicx}
\usepackage{txfonts}
\usepackage{subcaption}         

\usepackage{threeparttable}

\newcommand{\kms}{km~s$^{-1}$}

\begin{document}

 \title{Two young open clusters in Cygnus and their vicinity: combining
    multicolor photometry with \textit{\textbf{Gaia}} DR3 astrometry}

\author{S. Raudeli\={u}nas\inst{1}
     \and R. P. Boyle\inst{2}
     \and R. Janusz\inst{3}
     \and J. Zdanavi\v{c}ius\inst{1}
     \and M. Maskoli\={u}nas\inst{1}
     \and D. Semionov\inst{1}
     \and K.~\v{C}ernis\inst{1}
     \and V. \v{C}epas\inst{1}
     \and A. Kazlauskas\inst{1}
}
\institute{Astronomical Observatory of Vilnius
    University, Saul\.etekio al. 3, Vilnius LT-10257, Lithuania\\
    \and Vatican Observatory Research Group, Steward Observatory,
    Tucson, AZ 85721, USA\\
    \and Vatican Observatory, V-00120, Vatican City State
     }

\date{Received January 25, 2025}


  \abstract
 {Studies of young open clusters that reside mostly in heavily obscured environments with patchy distribution of
    absorbing material, such as the one in Cygnus, necessitate a combination of {\it Gaia} data with external deep
    multicolor photometry capable of accounting correctly for differential extinction.}
 {We investigate two neighboring clusters in the Cygnus complex, Berkeley 86 and Berkeley 87, with a primary emphasis
    on the evaluation of extinction in the field of view towards and across the clusters. We also analyze their kinematic
    behavior in space and time to discern their possible common origin and relation to the Cyg~OB1 association.}
 {New CCD photometry in the {\it Vilnius} seven-color system, obtained down to $V$\,=\,19.0~mag in the fields of these two
    clusters, is used to classify stars in terms of spectral and luminosity classes and to determine
    the individual values of interstellar extinction. The probable cluster members are identified in a 5-parameter
    space based on {\it Gaia} DR3. The cluster ages and stellar masses are derived through the use of the HR diagrams.
    To obtain the 3D kinematics of the clusters and trace their orbits back in time, we combine the {\it Gaia}-based proper-motions and distances with radial velocities from the literature.}
 {The estimated cluster properties show that both clusters are almost equidistant (1.7 kpc) and nearly coeval, with
 average ages of 6.1$\pm$0.5 and 6.5$\pm$0.4 Myr, respectively, and age dispersion of 3 Myr. The nonuniformity of extinction is evident within each cluster, especially pronounced across the face
of Berkeley 86 where the most-massive stars show substantial substructure. By extrapolating the observed mass
function to a minimum stellar mass, we obtain cluster masses of 519 $\mathrm{M}_\sun$ and 1551 $\mathrm{M}_\sun$
for Berkeley 86 and 87, respectively. Although both clusters share very similar properties, their orbital paths
show no indication that they had a common birthplace, however Berkeley 87 and its neighbor NGC\,6913 are very
likely to have been born in pair.}
 {}

\titlerunning{Two young opencluster in Cygnus}
  \maketitle

\section{Introduction}

After careful review of the Palomar Sky Survey plates, \citet{Setteducati1962} detected new open star clusters
(OCs) which now are known as the Berkeley clusters. Two of them, \object{Berkeley 86} and \object{Berkeley 87}, are located in the
Great Cygnus Rift region and are attributed to the \object{Cyg OB1} association along with the neighboring OCs \object{NGC 6913} and
\object{IC 4996} \citep{1978ApJS...38..309H,1992AJ....103..916F}.

Berkeley 86 (RA$_{2000}$\,=\,20:20:20.2, DEC$_{2000}$\,=\,+38:41:17; $\ell$\,=\,76$\degr.65$,
$b$\,=\,+1$\degr.28$) is a sparsely populated cluster. It lies within a cavity in the CO gas distribution
\citep[see, e.g., CO maps in][]{2024AJ....167..220Z} that could have been formed by powerful stellar winds from
the surrounding young stars of the Cyg OB1 association. The cluster contains very few bright stars, one of which,
\object{HD 228989}, is known as a spectroscopic binary of type O9.5\,V+O9.5\,V \citep{1995ApJ...454..151M}.

Berkeley 87 (RA$_{2000}$\,=\,20:21:43.0, DEC$_{2000}$\,=\,+37:22:14; $\ell$\,=\,75\degr.72, $b$\,=\,+0$\degr.30$)
is a relatively dense cluster, located at the southwestern edge of the giant star-forming complex Cygnus-X. The
cluster is most notable for hosting the rare oxygen-type Wolf-Rayet star \object{WR 142}, also known as ST3 \citep{1966AJ.....71..477S}
and \object{Sand 5} \citep{1971ApJ...164L..71S}, and the red supergiant star \object{BC Cyg}, both representing the evolutionary
stage before a supernova explosion. Berkeley 87 also contains the emission-line blue supergiant \object{HD 229059} which is the
brightest star in the cluster and a few other emission-line B-type stars.
\begin{figure}
   \centering
   \includegraphics[width=\hsize]{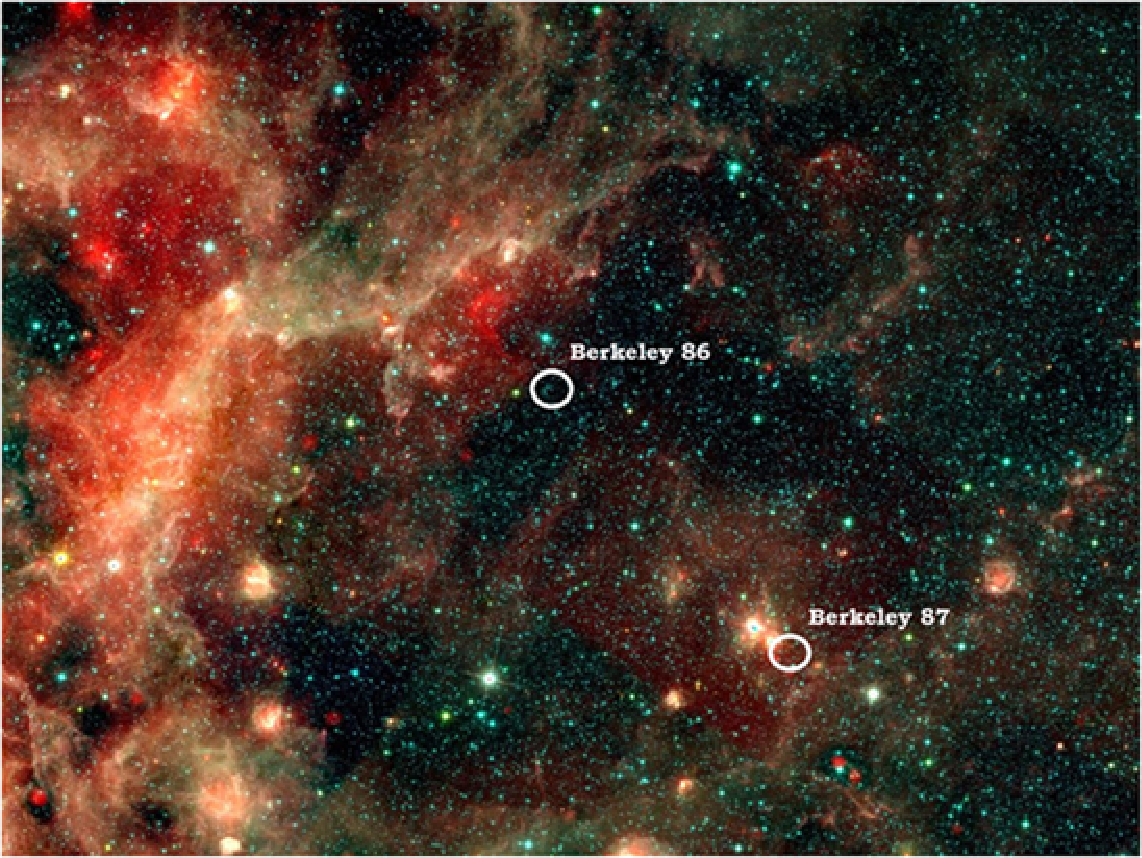}
      \caption{Location of Berkeley 86 and 87 in Cygnus (AllWISE color image in the background).} \label{fig:AllWISE}
  \end{figure}

The first studies of Berkeley 86 were carried out by \citet{1981PASP...93..441F} and \citet{1992AJ....103..916F},
and those of Berkeley 87 by \citet{1982PASP...94..789T}, who provided the estimates of distances, ages, and
reddenings from photometric and spectroscopic observations of the brightest stars in the clusters.
\citet{1995ApJ...454..151M,2001AJ....121.1050M} have determined MK spectral types for a number of stars in both
clusters, in addition to those available from the studies cited above. In later years, extensive astrometric
catalogs such as Hipparcos/Tycho-2, ASCC-2.5, PPMXL, UCAC), as well as other available data sources, were employed
to provide mean distances, proper motions, and other characteristics for Galactic OCs, including Berkeley 86 and
87 (e.g., a series of OCs catalogs by
\citet{2002A&A...389..871D,2006A&A...446..949D,2018MNRAS.478.5184D,2019MNRAS.486.5726D},
\citet{2005A&A...438.1163K,2013A&A...558A..53K,2016A&A...585A.101K,2023A&A...672A.187J}, among others). With the
release of the {\it Gaia} data, a number of recent works have resulted in all-sky catalogs of OCs, based on DR2
(\citet{2018A&A...618A..93C,2020A&A...640A...1C}; \citet{2019MNRAS.486.5726D,2021MNRAS.504..356D};
\citet{2019ApJS..245...32L}; \citet{2021ApJ...923..129J}) and EDR3/DR3 (\citet{2021A&A...651A.104P};
\citet{2024A&A...686A..42H}), that contain greatly improved membership information and basic parameters of the
two Berkeley clusters. We summarize the published data on both clusters in Appendix Tables \ref{table:A1} and
\ref{table:A2}.

While the {\it Gaia}-based estimates of the distances to Berkeley 86 and 87 
are in good agreement, the age and extinction are not known with certainty. Most of the published estimates of age
are in the range 4--11 Myrs for Berkeley 86 and 5--8 Myrs for Berkeley 87, if to consider the literature sources
that give the distance estimates in accord with most recent values ($\approx$\,1.7 kpc). The values of extinction
for Berkeley 86 are in the range $A_V$\,=\,1.6--2.9 mag (when two extremely discrepant estimates, 0.7 mag and 3.5
mag are ignored). For the more obscured Berkeley 87, the extinction determinations vary from 3.5 mag to 5.2 mag. A
wide range of reddening and $A_V$ values is likely the result of a combination of reasons including non-member
contamination, differential reddening and possible extinction anomalies. Also, there is an obvious absence of
self-consistent results on the radial velocities (RVs) what inevitably complicates the calculation of 3D
kinematics. In general, most of the recent works cited in Tables A.1 and A.2 were global OC studies, not aimed at
a detailed analysis of each cluster.

In the present paper we investigate the two clusters on the basis of multicolor photometry and the {\it Gaia} DR3
astrometry. For multicolor observations, we used the {\it Vilnius} seven color photometric system with passbands
at 345($U$), 374($P$),405($X$), 466($Y$), 516($Z$), 544($V$) and 656($S$) nm, described in
\citet{1992msp..book.....S}. This system is able to derive both two-dimensional spectral types in
the MK system and color excesses due to interstellar reddening. The {\it Vilnius} system was used in earlier works of our
group to investigate other young clusters in Cygnus, NGC\,6913 \citep{2013BaltA..22..181M,2014AJ....148...89S}, IC\,4996 \citep{2019A&A...623A..22S}, and IC\,1369 \citep{2020AJ....159...95S}.

In Section 2 we describe our CCD observations in the {\it Vilnius} system and the data reduction
(Sect.\,\ref{subsec:CCDdata}), and present spectral classification of stars in the observed fields
(Sect.\,\ref{subsec:Photclass}). In Section 3 we investigate the interstellar extinction toward the clusters,
including tests of the extinction law within the areas of study (Sect.\,\ref{subsec:extViln}) and an additional
means of tracing extinction through the use of 2MASS and WISE photometry (Sect.\,\ref{subsec:ext2MASS}). In
Section \ref{sec:clustpar} we provide the {\it Gaia}-based membership probabilities and distances
(Sect.\,\ref{subsec:memb}), the mean extinction and its variations within the clusters
(Sect.\,\ref{subsec:extclust}), and present CMDs and HRDs which are used to infer the cluster ages and stellar
masses (Sect.\,\ref{subsec:CMD}). The kinematics of the clusters is described in Section 5, where we derive the
peculiar 3D velocities (Sect.\,\ref{subsec:pecvel}) and galactic orbits (Sect.\,\ref{subsec:traceback}). In
Section \ref{subsec:CygOB1} we discuss the relation to the Cyg OB1 association. The conclusions are summarized in
Section\,\ref{sec:concl}.

\section{Photometry in the {\it Vilnius} system}
\subsection{CCD observations and data reductions} \label{subsec:CCDdata}

The observations of both clusters were obtained with the Vatican Advanced Technology Telescope (VATT) on Mt.
Graham, Arizona, and its 4k-CCD camera producing an image scale of  $0.37^{\prime\prime}$ pixel$^{-1}$ and a sky
coverage of $12'\times 12'$. The first run of observations started in April 2011, using filters of the {\it
StromVil} photometric system. The short exposures showed promising results, therefore deeper exposures
(2012–2014 and 2020–2021) were conducted with VATT, using the {\it Vilnius} photometric system. In April 2018,
the Maksutov-type 35/51 cm telescope at the Moletai Observatory in Lithuania was used to improve the tie-in of the
Berkeley 86/87 clusters to M\,29 (NGC\,6913) and to check the standards obtained for calibration. To ensure
accurate determination of color terms for the transformations, as well as of flat-field corrections, the standard
fields of the OCs M\,67 and/or M\,29  were observed during each run. The transformation equations between the
instrumental and standard systems for $V$ magnitudes and six color indices are
\begin{equation}V = v + c_{V}(y - v) + c_{0V}\end{equation}
\begin{equation}Y - V = y - v + c_{YV}(y - v) + c_{0YV}, etc.,\end{equation}
where $V$, $Y$,... denote the magnitudes in the standard {\it Vilnius} system and  $v$, $y$,... refer to
instrumental magnitudes; $c$ and $c_0$ are the color term coefficients and constants, respectively.

The errors of transformation to the standard system were of the order of 0.02 mag. Taking into account the
measurement errors, the typical uncertainties of final standard values of color indices were 0.035 mag for
$U$--$V$ and $\leq$\,0.030 mag for the remaining five color indices, while the mean errors for  $V$ were nearly
all within 0.020 mag.

The catalogs of photometry in the standard {\it Vilnius} system contain 1053 stars down to $V$$\approx$19 mag in
the area of Berkeley 86 (Table\,\ref{table:1}) and 502 stars in the area of Berkeley 87 (Table\,\ref{table:2}),
also down to $V$$\approx$19 mag. A few percent of stars in the Berkeley 86 field and as many  as one third of stars
in the Berkeley 87 field are very faint in the ultraviolet range and their magnitudes  $U$ and $P$ could not be
measured to sufficient accuracy; in such cases, their color indices  $U$--$V$ and $P$--$V$ are not given. Such a
large difference in the numbers of stars without photometric measures in the ultraviolet is obviously a
consequence of the much heavier extinction toward Berkeley 87.

\subsection{Photometric classification} \label{subsec:Photclass}

For two-dimensional classification of stars, the updated \mbox{QCOMPAR} code, described in our previous publications
\citep[e.g.][]{2020AJ....159...95S}, was applied.

As a measure of quality of the classification, the following quantity is used
\begin{equation}
    \sigma_Q = \sqrt{{\sum_{n}^{} \Delta Q_i^2}\over n},
\end{equation}
where $\Delta Q$ are differences in the corresponding $Q$-parameters of the program star and the standard star,
and $n$ is the number of the $Q$-parameters involved.

Our experience shows that the quality of classification $\sigma_Q$\,$\leq$\,0.05 mag is sufficient for the
determination of spectral and luminosity classes and interstellar extinction $A_V$ to an acceptable level of
accuracy.

The results of classification are given in Table\,\ref{table:1} (the Berkeley 86 area) and Table\,\ref{table:2}
(the Berkeley 87 area), along with photometry in the {\it Vilnius} system. The tables contain the magnitude $V$,
six color indices with their uncertainties, and photometric spectral types in the MK system.  The uncertainties
take into account the measurement errors and the errors of transformation to the standard system which are of the
order  of 0.02 mag. For stars too faint to be measured in the ultraviolet, the color indices $U$--$V$ and $P$--$V$
are not listed. We do not present spectral types when the classification quality $\sigma_Q$ exceeds 0.2 mag. When
$\sigma_Q$ is in the range 0.1--0.15 mag, the spectral type is marked as uncertain (using a colon), and when in
the range 0.15-0.2 mag, as doubtful (with a question mark).

\section{Interstellar extinction in the directions of Berkeley 86 and 87}
\subsection{Results from {\it Vilnius} photometry} \label{subsec:extViln}

To determine the interstellar extinction properly, we need to know the ratio of  total to selective extinction
($R_V$) toward both clusters, since in the literature some authors admit possible regional variations in the
interstellar extinction law in Cygnus. In the direction of Berkeley 86, \citet{2012A&A...541A..95F} estimated an
almost normal extinction law, with $R_V$\,$\sim$\,2.9 (see their Fig. 3). A similar value, $R_V$\,$\sim$\,2.8, was
found for Berkeley 87 in the recent work of \citet{2021A&A...650A.156D}. Most researchers agree that extinction in
the Berkeley 86 region is rather uniform. However, \citet{2001MNRAS.328..370Y} found it to be variable ($E(B-V)$
between 0.24 and 1.1 mag). \citet{2007BASI...35..383B} have found significant variations in $E(B-V)$ for Berkeley
87, which show a non-uniform distribution of interstellar material within the cluster. Also, the wide main
sequence in the CMD could be an indication of significant differential reddening across the cluster.

The regions of both clusters contain a number of hot B--A type stars which have $UBV$ and 2MASS observations.
Thus, there is a possibility of testing the $R_V$ values by means of the ratios of color excesses
$E_{V-J}/E_{B-V}$, $E_{V-H}/E_{B-V}$ and $E_{V-K}/E_{B-V}$, and the equations of \citet{1999PASP..111...63F}. For
this aim we selected 28 and 23 stars of B--A type in the regions of Berkeley 86 and Berkeley 87, respectively.
Their $J$, $H$, and $K$ magnitudes were taken from 2MASS and transformed to the standard $JHK$ system using the
equations derived by \citet{2001AJ....121.2851C}. The values of $V$ and $B-V$ were taken from the AAVSO APASS
survey\footnote{~http://www.aavso.org/apass} DR9 \citep{2015AAS...22533616H}. We used the intrinsic indices
$(V-J)_0$, $(V-H)_0$, $(V-K)_0$, and $(B-V)_0$ that are given in \citet{1992msp..book.....S}. The calculated $R_V$
values are 2.79 $\pm$ 0.16 for Berkeley 86 and 2.73 $\pm$ 0.15 for Berkeley 87. These results are similar to, but
slightly lower than, those given by \citet{2012A&A...541A..95F} and \citet{2021A&A...650A.156D}. In the {\it
Vilnius} system, the corresponding values are $R_{YV}$=3.74 for Berkeley 86 and  $R_{YV}$=3.66 for Berkeley 87.

The color excesses of stars are calculated by the equations:
\begin{equation}E_{Y-V} = (Y-V)_{\rm obs} - (Y-V)_0, \end{equation}
where $(Y-V)_0$ are the intrinsic color indices taken from \citet{1992msp..book.....S}. The values of $E_{Y-V}$
are transformed to the extinction $A_V$ by the equation
\begin{equation} A_V = R_{YV}\,E_{Y-V}.
\end{equation}
A typical uncertainty in $A_V$ is $\approxeq$ 0.15 mag \citep{2015AJ....149..161S}, due to the observational errors of $Y-V$ and the errors in
the intrinsic colors $(Y-V)_0$. The error in $(Y-V)_0$, and hence in extinction, depends mostly on uncertainty in
the derived spectral class, while the error of the luminosity class is much less important.

\begin{figure}
    \centering
    \includegraphics[width=\hsize,trim={2.5cm 2.2cm 2.5cm 2.2cm},clip]{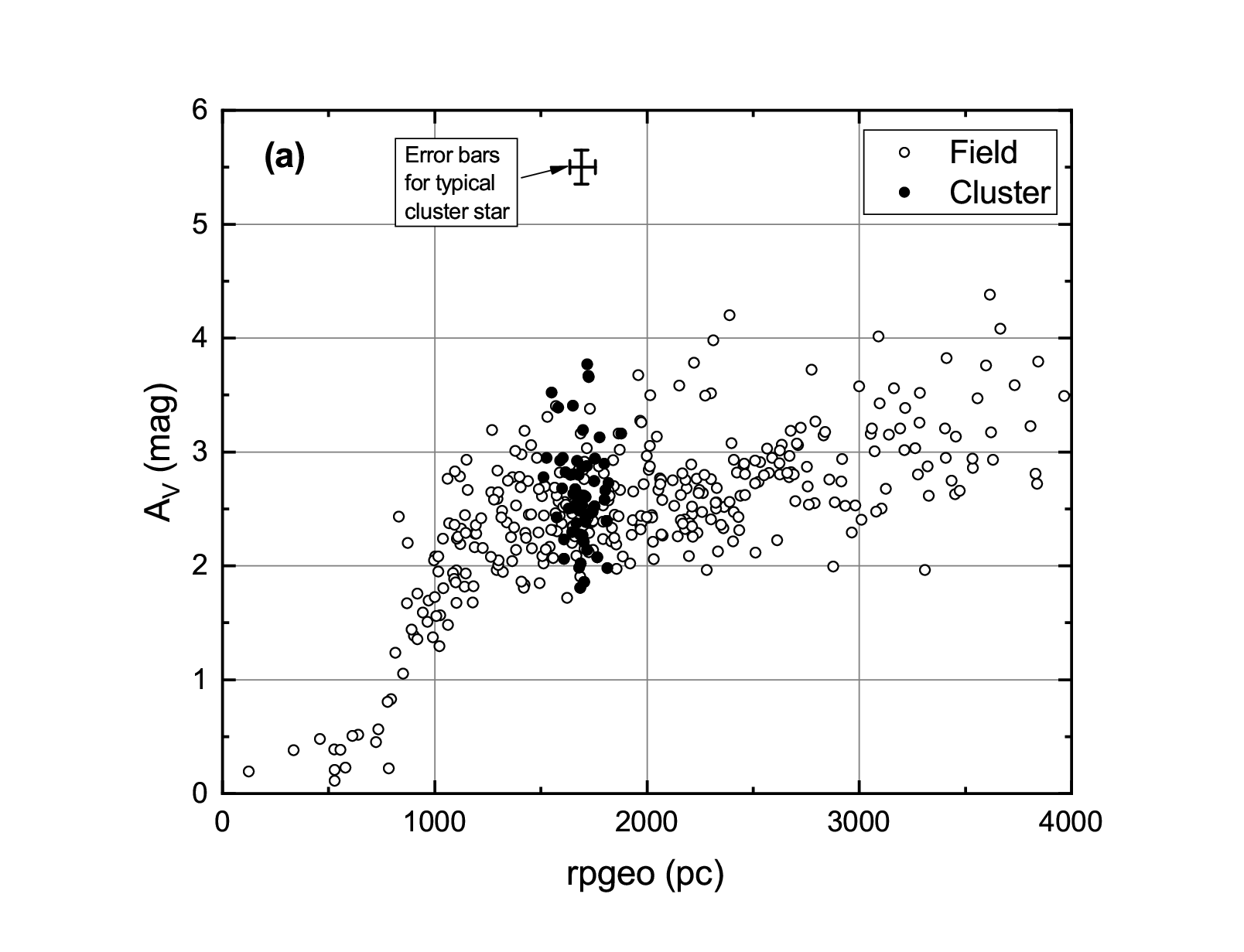}
    \includegraphics[width=\hsize,trim={2.5cm 1.0cm 2.5cm 2.2cm},clip]{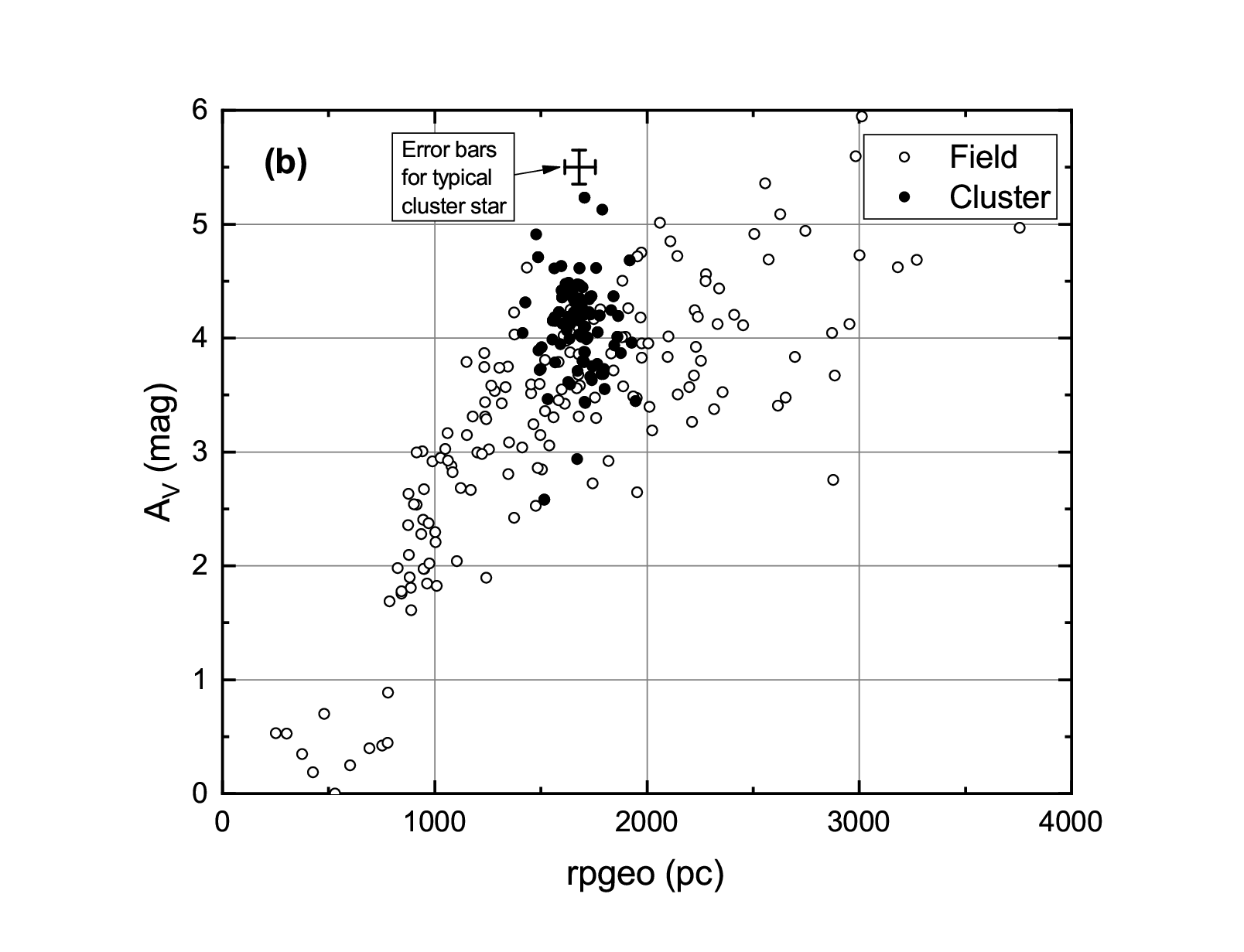}
    \caption{Interstellar extinction in the direction of Berkeley 86 (a) and Berkeley 87 (b). Black  dots show the
        positions of the identified cluster members (see Sect.\,\ref{subsec:extclust}).} \label{fig:Av-r}
\end{figure}

For the investigation of extinction, we used only those stars which have the classification errors $\sigma_Q\leq$
0.05 mag. Distances to the stars were taken from the catalog of \citet{2021AJ....161..147B}. We used the
photogeometric ($rpgeo$) distances from this catalog, where they were calculated from the {\it Gaia} EDR3
parallaxes converted to distances, taking into account nonlinearity of the transformation and asymmetry of the
resulting probability distribution. In the area of Berkeley 86, we have 758 stars with the classification quality
$\sigma_Q$ better than 0.05 mag, out of 1053 stars observed down to $V\approxeq$ 19 mag. The region of Berkeley 87
contains roughly two times fewer stars in the same sized $12'\times 12'$ field. There we observed 502 stars down
to $V\approxeq$ 19 mag, out of which 403 stars have our $A_V$ determinations. Plots of the interstellar extinction
$A_V$ versus distance for both clusters are shown in Fig.\,\ref{fig:Av-r}.

It is evident that in the direction of both clusters a steep rise of extinction, up to 2 mag, happens at 600--800
pc. Most probably this jump is related to the dust clouds of the Great Cygnus Rift, located at $\sim$700--900 pc
\citep{2016A&A...585A..31S}. However, the character of extinction at greater distances is different: while in the
direction of Berkeley 86 the growth of $A_V$ is quite moderate, up to 3.0--3.5 mag at 3--4 kpc, the extinction in
the direction of Berkeley 87 continues to rise up to $\sim$3 kpc where it reaches $\sim$ 4.5--5.0 mag.

\subsection{Results from 2MASS and WISE photometry of field  RCG stars} \label{subsec:ext2MASS}

Another way to investigate the interstellar extinction versus distance is to use red clump giant (RCG) stars which
have a low dispersion of their intrinsic colors and absolute magnitudes. Using the diagrams  $J-H$ vs. $H-K_S$, $H-W2$ vs. $J-W2$, and $K_S$ vs.
$H-K$ (here, $W2$ is the magnitude of the WISE system and the rest magnitudes are of the 2MASS system), we can
distinguish between RCGs and the other stars in the field. A detailed description of the method can be found in the paper by \citet{2016A&A...585A..31S}.

\begin{figure}
    \centering
    \includegraphics[width=\hsize,trim={2.5cm 1.2cm 2.5cm 2.2cm},clip]{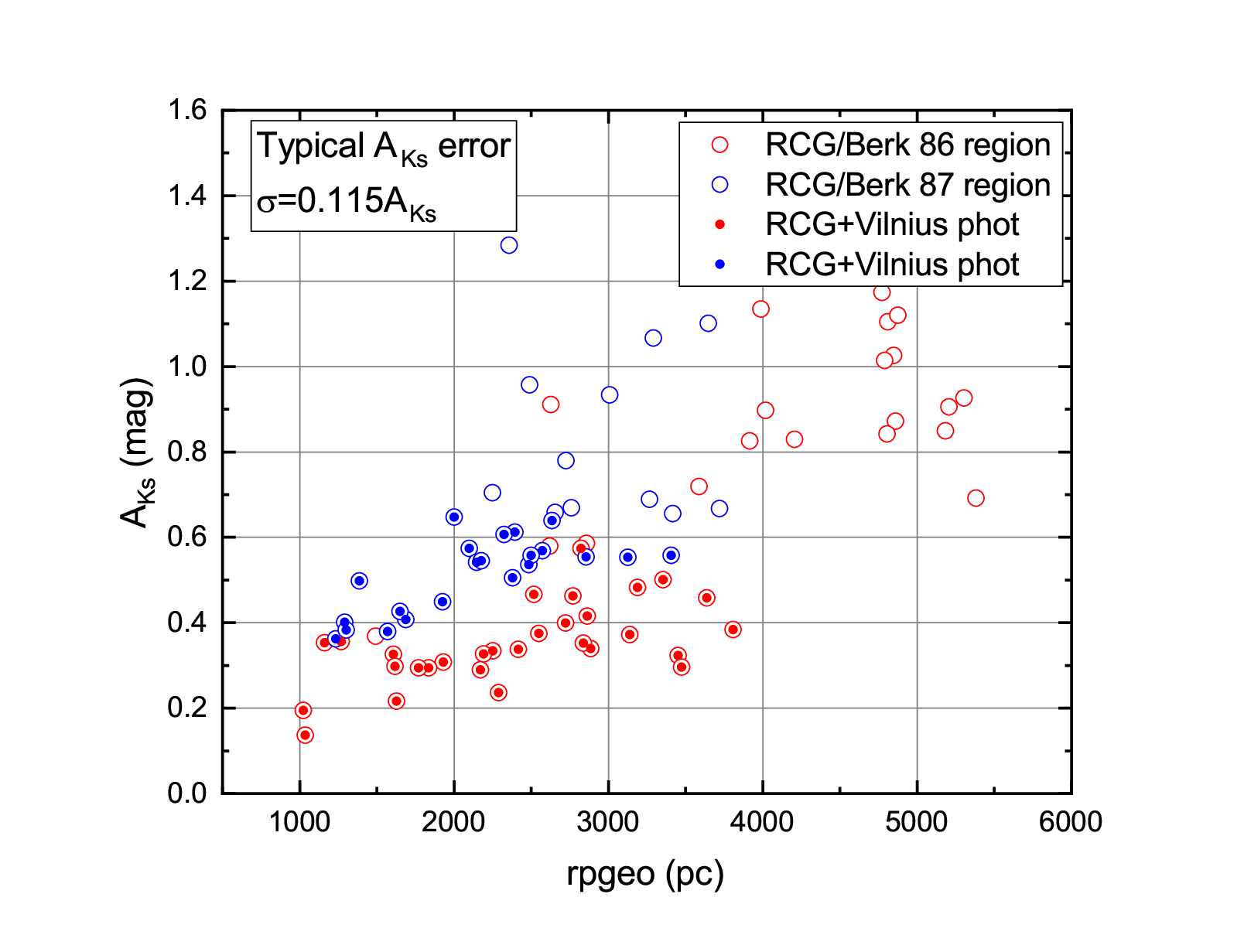}
    \caption{Extinction $A_{K_S}$ vs. distance for 49 RCGs in the direction of
        Berkeley 86 (red symbols) and 34 RCGs toward Berkeley 87 (blue symbols), selected from the 2MASS and WISE catalogs. Open circles denote all
        RCG stars in the fields, dots mark those observed in the {\it Vilnius} system.}
        \label{fig:Ks-r}

\end{figure}
Applying this method for the identification of  RCG candidates, we selected 58 suspected  RCG stars in the field
of Berkeley 86 and 40 such stars in the field of Berkeley 87. After excluding close binary stars (to avoid blended
measurements that can distort photometric and astrometric data), stars with negative parallaxes (which indicate
unreliable or erroneous distance estimates), and those without $rpgeo$ distance data, we are left with 49 stars in
the Berkeley 86 region (30 of which have {\it Vilnius} photometric measurements) and 34 stars in the Berkeley 87
region (22 of which have {\it Vilnius} measurements). These stars were dereddened by the method described in
\citet{2016A&A...585A..31S}, and their values of $A_{K_S}$ are plotted against distance in Fig.\,\ref{fig:Ks-r}.
Here we see that in the direction of Berkeley 86 the extinction $A_{K_S}$ can reach 1.1 mag at a distance of 4--5
kpc, what corresponds to $A_V$$\approx$ 9.1 mag. Up to 3.8 kpc in the direction of Berkeley 87, the RCG stars
selected by the same method show higher extinction than that toward Berkeley 86 in the same distance range. Beyond
3.8 kpc, however, we have no RCGs identified in the field of Berkeley 87. Comparing the interstellar extinction
toward each of the clusters at $\sim$2.5 kpc with that in Fig.\,\ref{fig:Av-r}, we see that the results from {\it
Vilnius} and 2MASS/WISE photometry agree reasonably well. We were not able to compare the trend of extinction at
distances $<$\,1 kpc, where we have very few RCGs identified because their images are saturated in the WISE $W2$
passband.

\section{Cluster parameters} \label{sec:clustpar}
\subsection{Membership and astrometric parameters} \label{subsec:memb}

Cluster membership probabilities $p$ were calculated using the extreme deconvolution Gaussian mixture program from
\citet{2009ApJ...700.1794B}. As a membership probability, the hyperparameter $q$ (their Eq. (16)) was used. We
performed clustering in a $12\arcmin\times12\arcmin$ area of each cluster, using five parameters from {\it Gaia}
DR3 for each star in our catalog of {\it Vilnius} photometry: equatorial coordinates, proper motions in right
ascension and declination, and parallax. We have found 63 stars with the membership probability $\geq$\,0.70 for
Berkeley 86 and 92 stars with the same probabilities for Berkeley 87. In fact, nearly all (95\%) of these stars
have probabilities $p$ higher than 0.9, with the rest few having $p$=0.7--09. The lists of these probable cluster
members are presented in Tables\,\ref{table:3} and \ref{table:4} for Berkeley 86 and 87, respectively.


\begin{figure}[]
    \centering
    \includegraphics[width=\hsize,trim={0.8cm 2.2cm 2.5cm 0.3cm},clip]{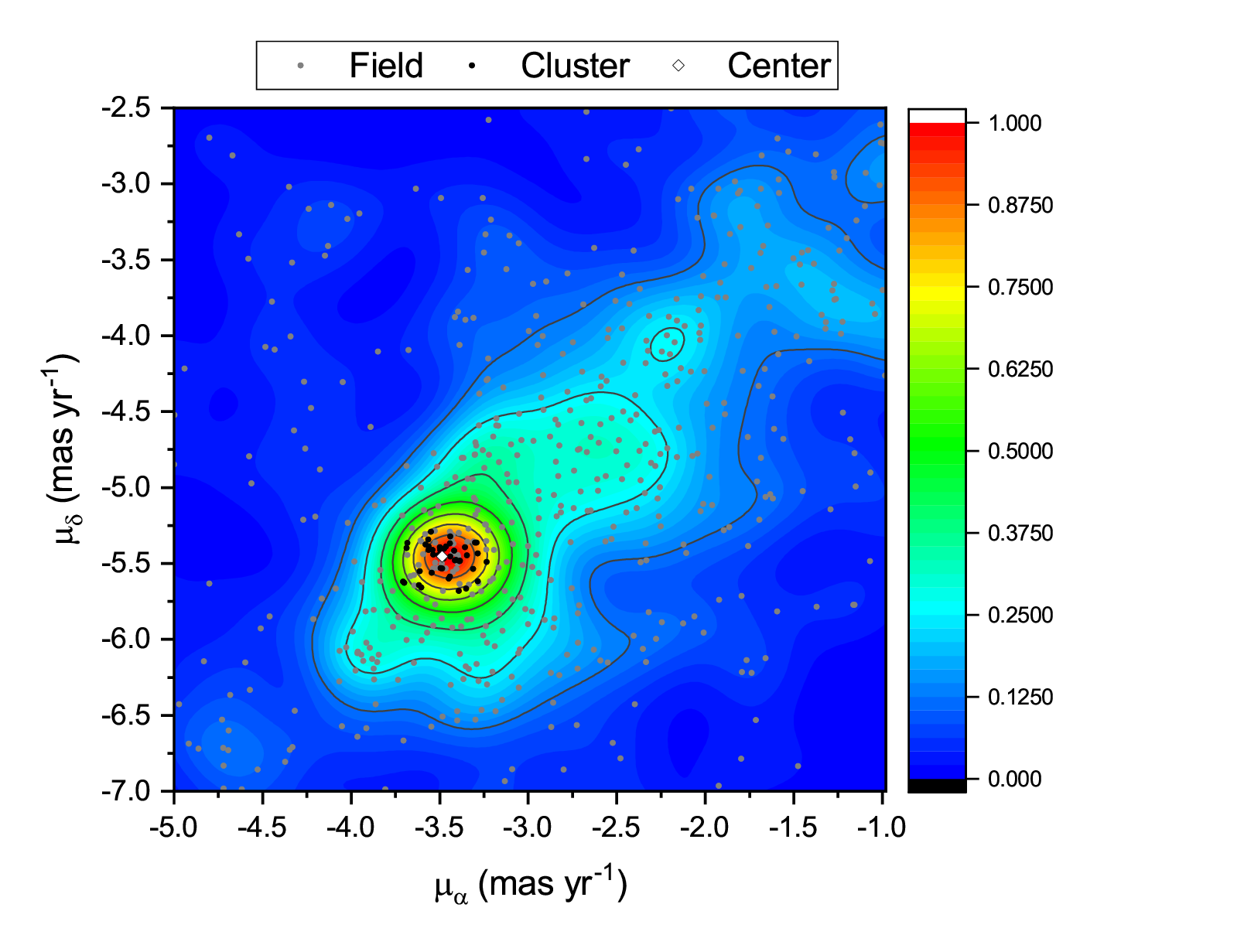}
    \includegraphics[width=\hsize,trim={0.8cm 0.6cm 2.5cm 2.2cm},clip]{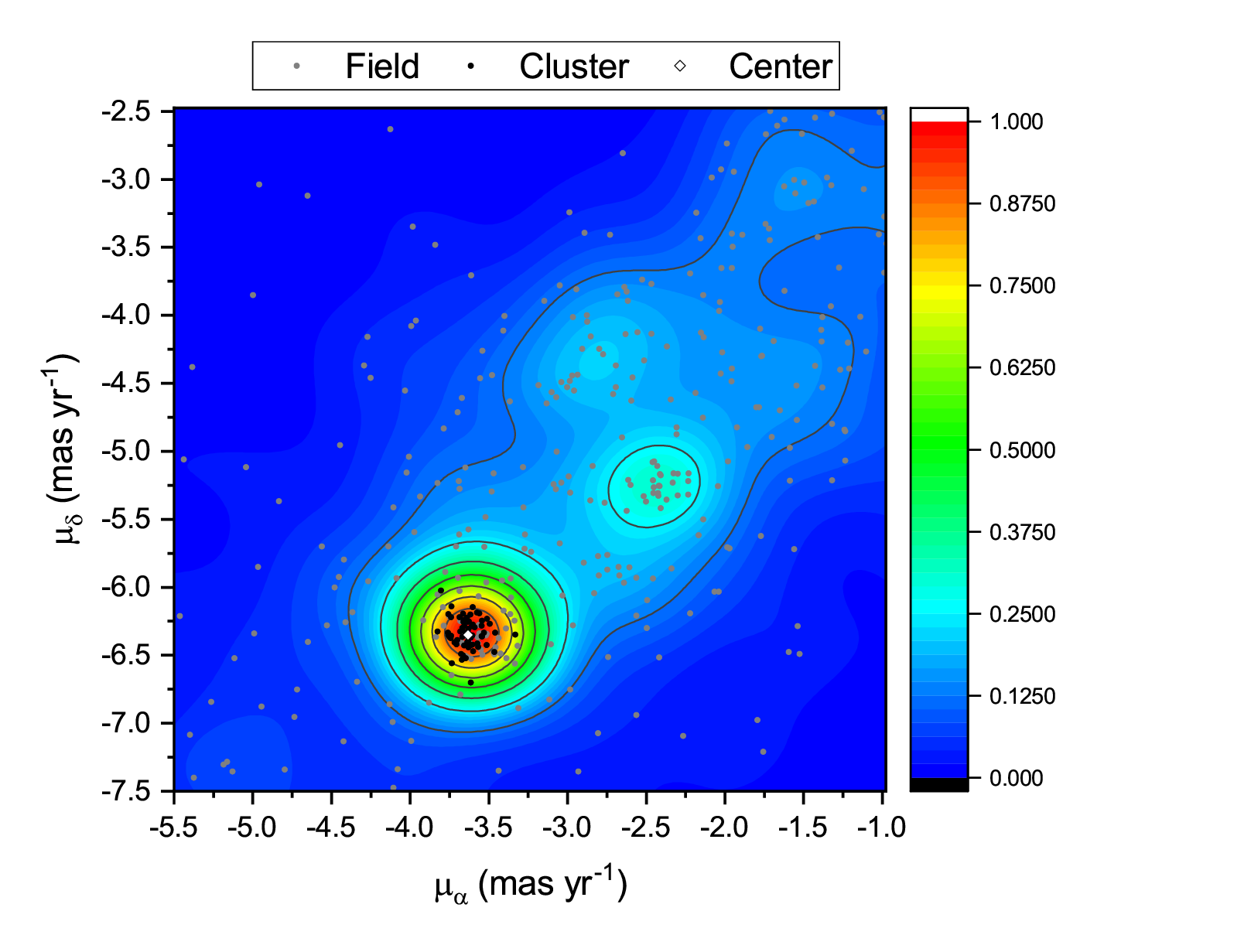}
    \caption{Contour maps in the proper motion plane ($\mu_{\alpha},\mu_{\delta}$) for Berkeley 86 (upper panel) and Berkeley 87 (lower panel).
        The scale of colors corresponds to the proper motion probability density normalized to 1.0 at the maximum value in the
        cluster's center. The grey dots mark the positions of field stars. The black dots are probable cluster stars. The white
        diamond denotes the mean value of proper motions of the cluster's center.} \label{fig:PMD}
\end{figure}

A comparison of our lists of member stars with those of \citet{2024A&A...686A..42H}, based also on the {\it Gaia}
DR3 data, has shown very good agreement between their probabilities and ours. However the numbers of the
identified members differ in that our lists contain more stars than theirs, despite the fact that our search for
probable members is restricted to a $12\arcmin\times12\arcmin$ field. In Berkeley 86, they identified 36 stars
with probabilities $p_{\rm HR}\geq$\,0.88, whereas our Table\,\ref{table:3} contains 59 stars with $p\geq$\,0.89.
Of those 36 members, we have 28 stars in common and all these appear in our member list; the remaining five are
absent in our list and three stars lie outside our field. In Berkeley 87, they presented a list of 82 stars with
$p_{\rm HR}\geq$\,0.98, including the well known red supergiant BC Cyg. Except for 13 stars lying outside our
field and five stars not included in our catalog, all of the remaining 64 stars are present in our member list. It
should be noted that the field of our CCD observations of Berkeley 87 is smaller than the entire size of this
cluster (see Fig.\,\ref{fig:mass}).

Fig.\,\ref{fig:PMD} illustrates the proper-motion probability density maps, thereby showing a narrow range in
$\mu_{\alpha}$ and $\mu_{\delta}$ possessed by probable cluster members ($p\geq$\,0.70). The medians of proper
motions are $\mu_{\alpha}$= --3.487$\pm$\,0.018, $\mu_{\delta}$ = --5.454$\pm\,$0.017 for Berkeley 86, and
$\mu_{\alpha}$ = --3.631$\pm\,$0.010, $\mu_{\delta}$ = --6.351$\pm\,$0.013 for Berkeley 87. These values are in
very good agreement with recent {\it Gaia} DR3 based results by \citet{2024A&A...686A..42H} (and
\citet{2021A&A...651A.104P} in the case of Berkeley  87), as well as with some earlier results from {\it Gaia} DR2
(e.g., \citet{2019MNRAS.486.5726D} and \citet{2019ApJS..245...32L}, in the case of Berkeley 86). The median
$rpgeo$ distances to the clusters, estimated from the same probable members, are 1691$\pm\,$9 pc for Berkeley 86
and 1681$\pm\,$9 pc for Berkeley 87. Here the proper motion and distance uncertainties are standard errors of the
mean, estimated from the median absolute deviation and multiplied by 1.4826. The corresponding true distance
moduli are 11.14 mag and 11.13 mag, respectively. At these distances to the clusters, the probable members in the
observed field of each of the clusters are all confined to within a radius of 5 pc.

\subsection{Extinction} \label{subsec:extclust}

\begin{figure}
    \centering
    \includegraphics[width=\hsize,trim={0.4cm 1.8cm 0.8cm 2.5cm},clip]{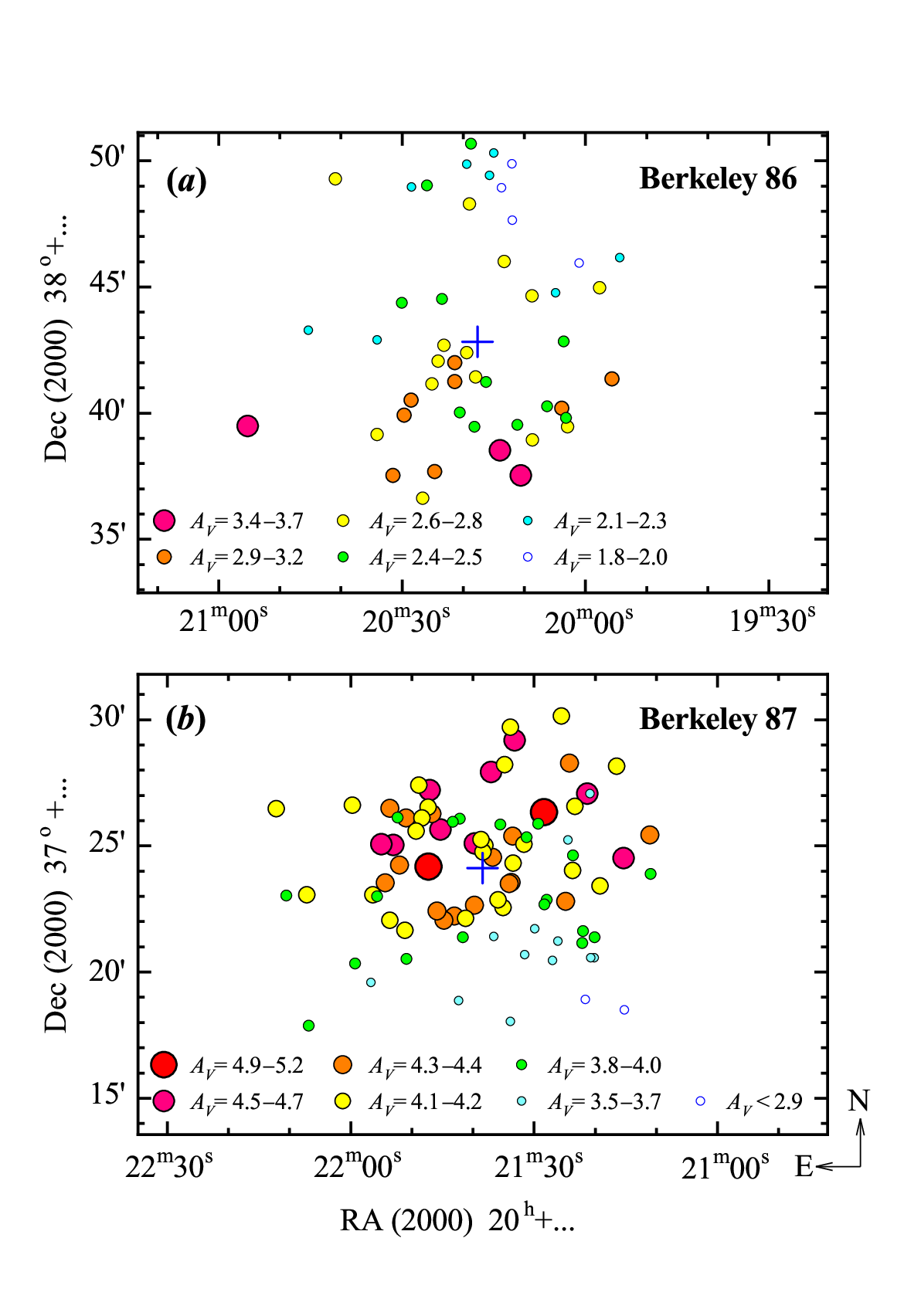}
    \caption{Extinction maps of the two clusters. Different size
        and colors of the circles correspond to different $A_V$ intervals indicated in the legends.
    }
    \label{fig:Av_map}
\end{figure}

In this and the following analysis we will use only those probable member stars ($p\geq$\,0.70) which have the
quality of photometric classification $\sigma_Q$ $\leq$\,0.05 mag. This restriction has removed about one-tenth of
stars from the lists of cluster members. Nearly all of the stars removed are of 18 mag or fainter. Thus our final
lists of member stars with good quality of classification and extinction determination contain 49 and 86 stars for
Berkeley 86 and 87, respectively.

Fig.\,\ref{fig:Av_map} shows a considerable scatter in the values of extinction for the member stars of both
clusters. In Berkeley 86, the $A_V$ values are distributed in the range 1.8--3.7 mag, with
$<$$A_V$$>$=2.57$\pm$0.38 mag. (Here, and below, we give a mean with a standard deviation.) This value agrees well
with the recent estimates of \citet{2021MNRAS.504..356D} and \citet{2024A&A...686A..42H}.

In Berkeley 87, the range of the individual values of extinction is somewhat wider, 2.6--5.2 mag, with
$<$$A_V$$>$=4.09$\pm$0.39. For this cluster, we compared our estimates of $A_V$ with the results of
\citet{2021A&A...650A.156D}. They have determined $A_V$ using the color excesses $E_{V-K_S}$ and $E_{J-K_S}$ for
11 member stars with MK types, nine of which have our $A_V$ estimates from {\it Vilnius} photometric
classification of good quality ($\sigma_Q$$<\,$0.05 mag). The range of their estimates of $A_V$ is 3.76--5.57 mag,
with a mean of 4.59$\pm$0.51 mag which is in agreement with our arithmetic average for the same nine stars
(4.35$\pm$0.42 mag), when the difference in the ratio of total-to-selective extinction used by these authors
($R_V$=2.8) and us ($R_V$=2.73) is taken into account. There are also other 13 B-type member stars in the catalog
of \citet{2021A&A...650A.156D}, the $A_V$ values of which are converted from $E_{B-V}$ given by
\citet{1982PASP...94..789T}. For this set of B-type stars, their mean ($<$$A_V$$>$\,=\,4.28$\pm$0.51) agrees
fairly well with that obtained from our photometry ($<$$A_V$$>$\,=\,4.23$\pm$0.25).

We emphasize that the spread in the $A_V$ values is significant enough that one could plausibly argue for the
presence of nonuniform distribution of the reddening in the areas of the clusters. Other factors, such as
unrecognized binarity, possible stellar variability or the presence of circumstellar dust, can also cause an
appreciable fraction of the scatter, but perhaps not as significant as to entirely account for the differences in
$A_V$ of the order of $\simeq$2 mag within the same cluster.

As the extinction maps in Fig.\,\ref{fig:Av_map} show there are some smaller or larger substructures across the
face of each cluster. The distinguishing feature of the distribution of stars in Berkeley 86 is the location of
the more-heavily reddened stars only to the south of the cluster's center. The median extinction in this
(southern) part of the cluster is 2.6$\pm$0.2 mag, as exhibited by 10 stars of spectral types B5 and earlier, or
2.7$\pm$0.1 mag, as obtained from the reddenings of the remaining 22 stars of later types (B9--F). To the north of
the center, where only the stars of types B9 and later are present (17 in number), the median extinction is lower,
2.3$\pm$0.1 mag. In the field of Berkeley 87, the largest values of extinction exhibit the member stars in the
central part, where they are clumped in patches with gaps between. On the south side of the cluster, the
interstellar extinction is evidently lower. Similar features of patchiness of the reddening distribution in the
Berkeley 87 area were first noted and described by \citet{1982PASP...94..789T}.

\subsection{HR diagrams, ages, and masses} \label{subsec:CMD}
\begin{figure}
    \centering
    \includegraphics[width=\hsize,trim={0.1cm 2.1cm 0.6cm 2.9cm},clip]{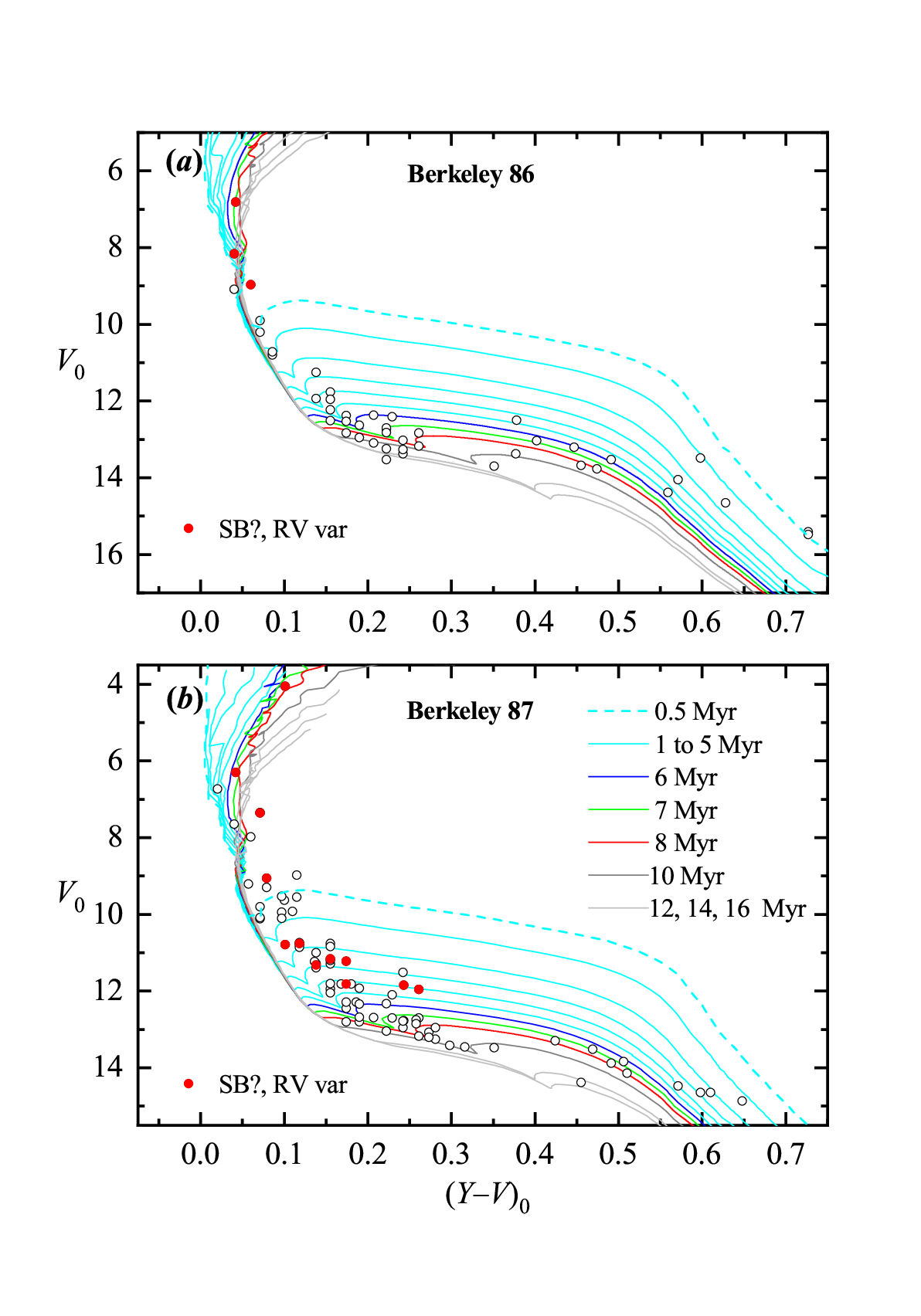}
    \caption{CMDs in the {\it Vilnius} system. Only stars with $\sigma_Q$$<\,$0.05 mag are plotted. Red open circles denote
        known RV variables and suspected SB stars. The $Z$\,=0.0152 Padova isochrones are shifted by
        distance moduli of $V_0-M_V$=\,11.14 mag and 11.13 mag for Berkeley\,86 and 87, respectively.}
    \label{fig:VYV}
\end{figure}
\begin{figure}
    \centering
    \includegraphics[width=\hsize,trim={0.1cm 2.1cm 0.6cm 2.9cm},clip]{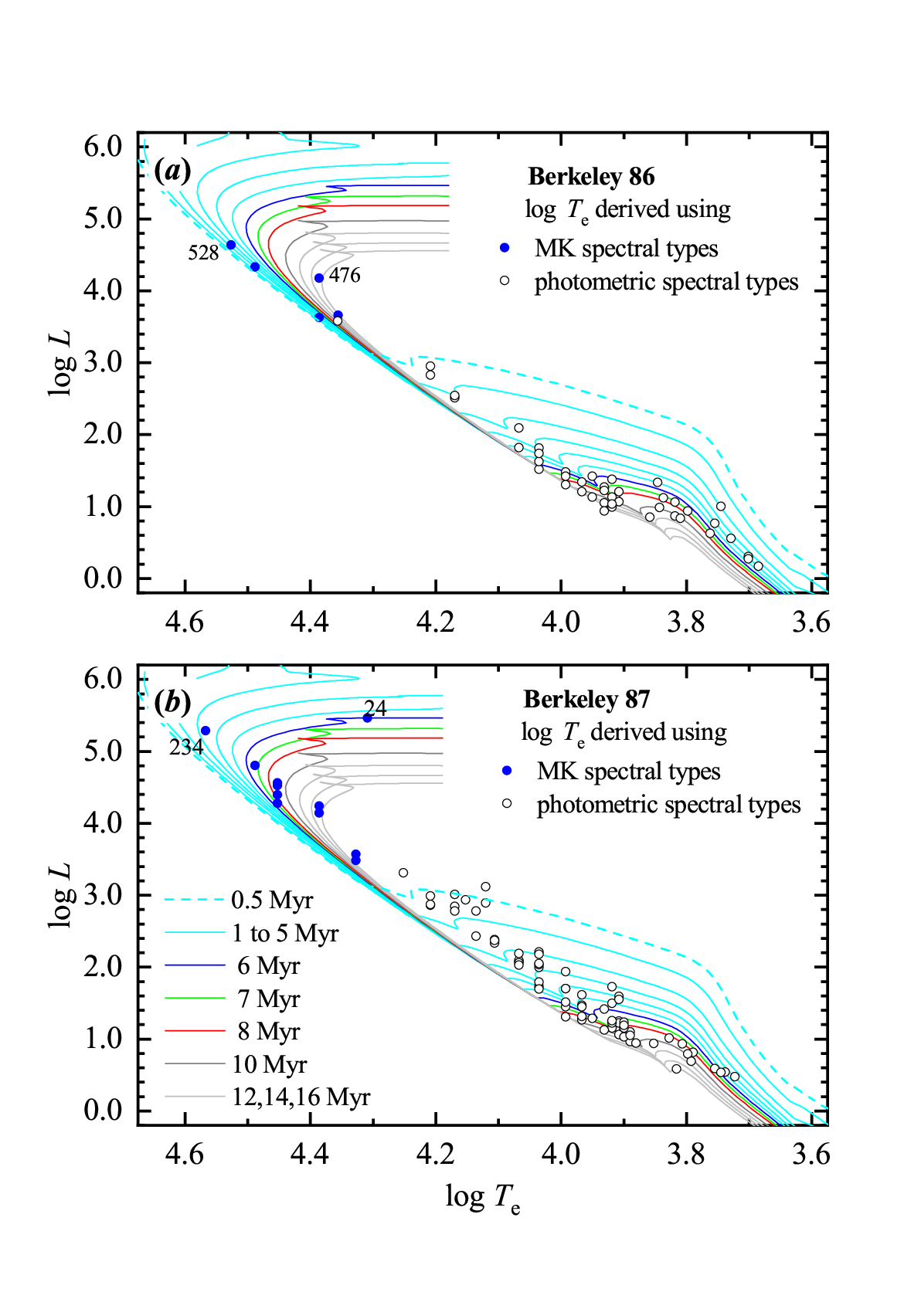}
    \caption{HR diagrams.
        The blue filled circles denote stars placed via their available MK types and the open circles
        represent the remaining stars placed via their photometry in the {\it Vilnius} system.
        The known SB star No.528 in Berkeley 86 is shown at the location of equal-mass components of type O9.5\,V.
    }
    \label{fig:HR1}
\end{figure}
The intrinsic color-magnitude diagrams $V_0$,\,$(Y-V)_0$ are shown in Fig\,\ref{fig:VYV}. Here, the magnitudes are
corrected individually for extinction $A_V$, and the colors $(Y-V)_0$ being taken from \citet{1992msp..book.....S}
according to spectral and luminosity class of the star. To derive stellar ages and masses, we use the Padova
database of stellar evolutionary tracks and isochrones for solar metallicity \citep{2002A&A...391..195G,
2012MNRAS.427..127B}, transformed to the observational plane of the {\it Vilnius} system. For both clusters, the
metallicities are not known with certainty, since very few estimates of [Fe/H] are available. The more recent ones
for Berkeley 86 are --0.08 dex of \citet{2003NewA....8..737T}, as derived from photometric data, and
0.18$\pm$0.308 dex of \citet{2021MNRAS.504..356D}. According to the latter source, the metallicity of Berkeley 87
is somewhat lower, [Fe/H]\,=\,--0.233$\pm$0.413. \citet{2023MNRAS.525.2315A} evaluated [Fe/H] to be
0.051$\pm$0.063 Berkeley 86 and 0.051$\pm$0.070 for Berkeley 87. As the cited values are nearly solar (subject to
significant errors), we consider the use of solar metallicity isochrones ($Z$\,=0.0152) to be quite acceptable.

Visual inspection of the panels in Fig.\,\ref{fig:VYV} shows that both clusters seem to be of similar age, with
their stars located mainly around the 6--8 Myr izochrones. In both diagrams we see numerous stars in the region to
the upper right of the main sequence (MS). Many of these can be considered as pre-main-sequence (PMS) stars. In
Fig.\ref{fig:VYV}(b), a group of stars at $(Y-V)_0$\,$\approx$\,0.05--0.1 (the B4--B6 star domain) is likely to
represent PMS stars of age less than 1 Myr, though unrecognized binarity or other factors can play a partial role.

At the upper part of each CMDs, we have very few stars to compare their location to the isochrones. Therefore in
age dating we included five B0.5--B1 stars with MK spectral types available in the literature (Table\,\ref{table:A1}), for
which our photometric classification was either uncertain (No.\,353 in Berkeley 86 and No.\,106 in Berkeley 87) or
their location on the CMD was not suitable for age determination (Nos.\,243, 298, and 314 in Berkeley 87). In the
case of the O9.5-type spectroscopic binary No.\,528 in Berkeley 86, we use the MK types of the components instead
of its photometric spectral type. For these few stars we give priority to using the theoretical HR diagram
(Fig.\,\ref{fig:HR1}). Other OB-type stars having MK spectral types were also involved in age dating, as a check
on the validity of our results obtained via photometry. By interpolating between the two side isochrones we
derived the individual ages and masses for nearly all member stars (when an unambiguous age determination was
possible).

In Berkeley 86, the most massive, O9.5--B5 stars (8--19 $\mathrm{M}_\sun$) lie on the MS, with the exception of
two stars for which the nearest isochrones are of 13 Myr and 15 Myr. In age dating, therefore, we could not rely
upon these only two stars. For nine most massive stars in Berkeley 87 (11--30 $\mathrm{M}_\sun$), we obtain an
average age of 8.9$\pm$1.3 Myr. However the intermediate mass stars (4--8 $\mathrm{M}_\sun$) in the latter cluster
all lie in the CMD/HRD diagrams between the 0.5 and 1 Myr isochrones; the age of 15 such stars can be formally
estimated as low as 0.8$\pm$0.1 Myr. For PMS stars of lower masses ($\leq3\mathrm{M}_\sun$), the average age is
the same for both clusters: 6.1$\pm$0.5 and 6.1$\pm$0.4 Myr, based on 39 and 60 stars in Berkeley 86 and 87,
respectively. In the case of Berkeley 87, its average age can be slightly higher, 6.5$\pm$0.4 Myr, if we take into
consideration also the high-mass stars at, and above, the MS turn point, for which age dating is less spurious
than that of the group of intermediate mass stars (4--8 $\mathrm{M}_\sun$).

The PMS population in each of the two clusters shows a skewed age distribution, with a spread even greater than
the mean cluster age (see the histograms in Fig.\,\ref{fig:hist_age1} in Appendix\,\ref{sec:appB}). The majority
of the stars cover a range of 0.6--14 Myr in Berkeley 86, and 0.3--15 Myr in  Berkeley 87. The age spread can in
part be accounted for by uncertainties in the two stellar parameters used in age dating. The separation in $V$
between the isochrones covered by the member stars is often less than a shift to be caused by stellar duplicity,
what would underestimate the stellar age by at most 2 Myr. Other factors, such as misclassification in luminosity
class, the errors in $A_V$ (up to 0.2 mag), stellar variability, the presence of circumstellar material and other
peculiarities, may contribute additional scatter in age, mainly in both directions. All these factors alone, or
taken together with some uncertainties in both the model isochrones and their conversion to observational
coordinates, can hardly explain the entire range of ages. Therefore, a significant part of the observed age spread
may be due to the fact that star formation at lower masses might continue on a timescale of nearly the order of
the mean cluster age. The ages for the most massive stars in both clusters are confined to within narrower ranges,
what can indicate that their formation occurred on much shorter timescales.

\begin{figure}
   \centering
   \includegraphics[width=\hsize,trim={0.4cm 1.6cm 0.8cm 2.5cm},clip]{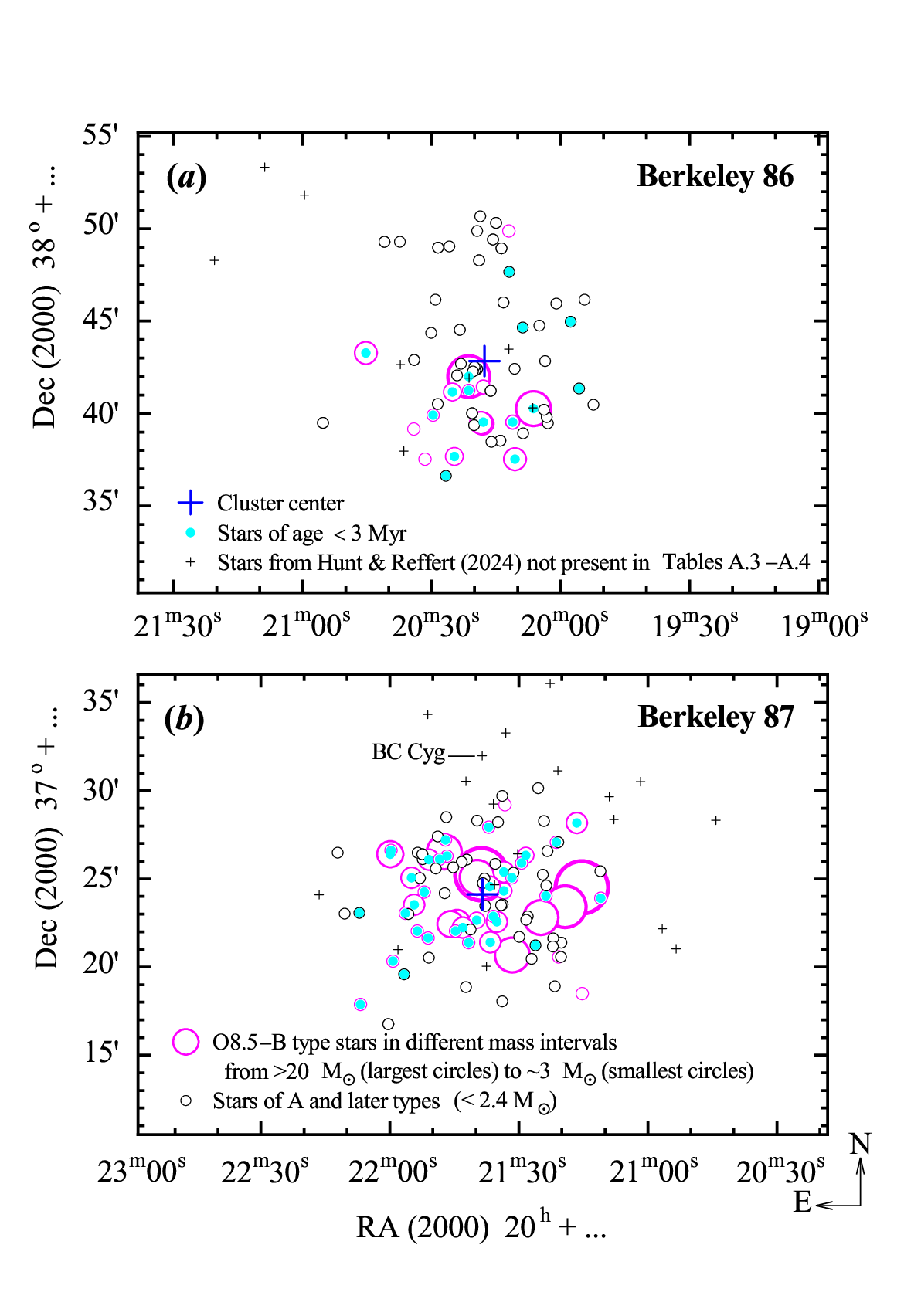}
      \caption{Maps of the distribution of member stars by mass. Different sizes
      of the circles in color are proportional to different mass range of O8.5--B stars: from >20\,$M_\sun$ (largest circles)
      to $\sim$3\,$M_\sun$ (smallest circles). Black circles represent stars of <2.4\,$M_\sun$ (early A- and later-type stars).
      Small cyan points superposed upon the symbols mark the youngest stars ($<$\,3 Myr). The stars
      with high membership probability ($p\geq0.88$) from \citet{2024A&A...686A..42H}, not present in our catalog of
      member stars (Tables\,\ref{table:3}--\ref{table:4}), are plotted as small cross symbols.
      } \label{fig:mass}
\end{figure}

Fig.\,\ref{fig:mass} demonstrates how the member stars of different mass are distributed across the area of each
cluster. The individual stellar masses are inferred from the isochrones, except for the star WR\/142 (No.\,285) in
Berkeley 87. The mass of this particular star, 11.9 M$_{\sun}$, has been derived using the value of $\log L$ from
\citet{2015A&A...581A.110T}, rescaled to the cluster's distance modulus 11.13 mag, and applying the $M/L$ relation
for WC/WO stars from \citet{1989A&A...210...93L}. In the figure we also plotted additional members identified by
\citet{2024A&A...686A..42H} that are not present in our catalog, just for comparison of our observational coverage
of the cluster areas.

In Berkeley 86, the high-mass stars show substantial substructure in their position: they all are located to the
south of the cluster center, where are concentrated stars having heavier reddening (see Fig.\,\ref{fig:Av_map}).
The median ages of the lower-mass stars ($\leq$\,3 $M_\sun)$ in both the northern and southern parts of the
cluster slightly differ, 7.2$\pm$0.8 Myr (14 stars) and 5.6$\pm$0.7 Myr (25 stars), respectively. Such
distribution of stars across the face of Berkeley 86 suggests that its southern area might have been the site of
later (secondary) stage of star formation within this cluster. The location of three member stars from
\citet{2024A&A...686A..42H} in the upper left corner of Fig.\,\ref{fig:mass}{\emph{\it a}} (at 0.23\degr--
0.28\degr, or 7--8 pc, from the center) may indicate a leakage of stars from the cluster.

In Berkeley 87 (Fig.\,\ref{fig:mass}{\emph{\it b}}), the distribution of high-mass stars also closely resembles
that of extinction shown in Fig.\,\ref{fig:Av_map}. Their central and eastern groupings both appear to be younger
($<$\,3 Myr) than the western group of a few massive stars of ages in the range 6--9 Myr. Fot this cluster, the
member list by \citet{2024A&A...686A..42H} contains 18 stars in addition to our list, 15 of which lie outside our
CCD field, including the red supergiant BC Cyg of 19 M$_\sun$ (Comer\'{o}n et al. 2020), located away from the
cluster nucleus and concentration of high-mass stars (hence this star is often considered to be a member of
Cyg~OB1, e.g. in the list by \citet{2020MNRAS.493.2339M}). According to the position of those additional stars in
the {\it Gaia}-based $(G,BP-RP)$ diagram, most of them could be of A and later spectral types, i.e., of mass
$\lesssim$\,3 M$_\sun$, and very few of higher mass (the cited catalog gives masses only for two of those 18
stars, and both are of $\leq$\,1.7 M$_\sun$). There are also an additional three or four B-type candidates outside
our field, the mass of which could be above 3 M$_\sun$, but they are in minority and do not change the general
picture seen in Fig.\,\ref{fig:mass}{\emph{\it b}} that the outer regions are dominated by low-mass stars. Thus,
there is some indication that either the more massive stars might have been formed preferentially near the center
or mass segregation has taken place in this cluster.

Finally, we use the individual stellar masses to derive the cluster mass by extrapolating the observed section of
its mass function (MF) to the full MF, assuming this section of stars is complete. In the case of Berkeley 86, a
reasonably large observational coverage may be expected down to $V$\,=\,16 mag. For Berkeley 87, however, the
field of our CCD observations does not encompass the entire area of this cluster (see
Fig.\,\ref{fig:mass}{\emph{\it b}}). In this case only the brighter end of the observed MF can be safely used,
assuming that there are no high-mass stars outside the field (though one massive member star, BC Cyg, lies outside
the central area). After fixing the maximum stellar mass and choosing carefully the minimum stellar mass at which
the number of observed stars is most complete, we calculated directly the sum of the individual masses in that
mass interval and used this value to obtain the normalization of the stellar MF (details are given in
Appendix\,\ref{sec:appB}). The cluster mass was then calculated by a multiple power-law of the Kroupa IMF
\citep{2004MNRAS.348..187W} for a mass range from the maximum stellar mass derived in the cluster to the 0.08
$\mathrm{M}_\sun$ limit.

We have not taken into account unrecognized binarity, except for the spectroscopic binary No.\,528 in Berkeley 86.
This star is likely to be the most massive in this cluster, with its components of $\sim$19 $\mathrm{M}_\sun$
each. We obtained the total mass of Berkeley 86 to be 519 M$_\sun$, or somewhat higher (633 M$_\sun$) when only
stars of mass $\geq$\,4 M$_\sun$ were counted. For Berkeley 87, with its star No.\,234 of 31 $M_\sun$ taken as
representing the maximum stellar mass in the cluster, the total mass would be 1551 $\mathrm{M}_\sun$, or higher if
binarity of star No.\,234 \citep{2004AN....325..380N} were taken into account. Here the member star BC Cyg was
admitted to our high-mass sample for the sake of completeness. The obtained values of cluster masses can be
considered as lower limits, since we made no attempt to correct the observed MF for binarity effects and data
incompleteness; evolutionary effects were also unaccounted for. We note, for comparison, that
\citet{2023MNRAS.525.2315A} applied their novel method based on the simulation of synthetic clusters to the {\it
Gaia} EDR3 data and obtained masses of 585$\pm$117 and 1779$\pm$355 M$_\sun$ for Berkeley 86 and 87, respectively.
The more recent estimates of their photometric masses (i.e., derived from stellar luminosity) are, however, lower:
317$\pm$64 and 1196$\pm$321 M$_\sun$ as based on {\it Gaia} DR3 \citep{2024A&A...686A..42H}, and 243$\pm$28 and
992$\pm$55 M$_\sun$ as based on {\it Gaia} DR2 data \citep{2025A&A...693A.305A}.

\section{Space motions}

In this section we use the mean values of proper motions and distances, derived in the previous section, and the
radial velocity (RV) data collected from the literature (Appendix\,\ref{sec:appC}) to obtain the three-dimensional (3D)
kinematics of the two Berkeley clusters and trace their orbits. Also, based on existing RV data and mean
astrometric parameters we calculate the kinematics of two other clusters of the Cyg\,OB1 association,
NGC\,6913\,(M~29) and IC\,4996, in order to compare the motions of all four clusters relative to one another over
their lifespan.

\subsection{Peculiar space velocities} \label{subsec:pecvel}

\begin{table*}
    \begin{threeparttable}
        \caption[]{Input data and kinematic parameters of four open clusters in
            Cygnus. The columns list, in succession: cluster age; distance; proper motion components and radial
            velocity; peculiar velocity components $U_\mathrm{s}$ (directed toward Galactic center), $V_\mathrm{s}$ (in the direction of
            Galactic rotation), and $W_\mathrm{s}$ (toward the north Galactic pole);
            distance $Z$ from the Galaxy's mid-plane at present time ($Z_{t=0}$);
            maximum $Z$ to be reached on the orbit; distance $Z$ at the time of birth according to a given age; time of crossing
            the mid-plane.}
        \label{uvw}
        \centering                         
         \begin{tabular}{l r c c c c r r r c c c r}
            \hline\hline \noalign{\smallskip}
            \multicolumn{1}{l}{Cluster} & \multicolumn{1}{r}{$Age$} & \multicolumn{1}{c}{$r$} & \multicolumn{1}{c}{$\mu_\alpha$}
            & \multicolumn{1}{c}{$\mu_\delta$} & \multicolumn{1}{c}{$v_r$} &
            \multicolumn{1}{c}{$U_\mathrm{s}$} & \multicolumn{1}{c}{$V_\mathrm{s}$} &
            \multicolumn{1}{c}{$W_\mathrm{s}$} & \multicolumn{1}{c}{$Z\,_{t=0}$} & \multicolumn{1}{c}{$Z\,_{\mathrm{max}}$}& \multicolumn{1}{c}{$Z\,_{\mathrm{birth}}$} & \multicolumn{1}{c}{$t_{Z=0}$}\\
            &(Myr)& \multicolumn{1}{c}{(pc)} & \multicolumn{1}{c}{(mas)} & \multicolumn{1}{c}{(mas)} &
            \multicolumn{1}{c}{(\kms\rlap{)}} & \multicolumn{1}{c}{(\kms\rlap{)}} & \multicolumn{1}{c}{(\kms\rlap{)}} & \multicolumn{1}{c}{(\kms\rlap{)}}
            &(pc)&(pc)&(pc)&(Myr)\\
            \hline \noalign{\smallskip} \noalign{\smallskip}
            Berkeley 86 & 6 & 1691 & --3.487 & --5.454 &  --13.6 &  10.9 & --6.0 & 5.2 & 58&85&15&--9.2\\
            \noalign{\smallskip}
            Berkeley 87 & 7 & 1681 & --3.631 & --6.351 & --17.4 &  17.0 & --10.9 &2.1 & 30&37&11&--10.6\\
            \noalign{\smallskip}
            NGC~6913 & 8.3\rlap{$^\ast$} & 1694 & --3.362 & --5.796 & --16.5 &  12.5 & --8.7 & 2.5 & 39&48&12&--11.6\\
            \noalign{\smallskip}
            IC~4996$^\ast$ & 10.6\rlap{$^\ast$} & 1954 & --2.636 & --5.351  & --11.5 &  5.9 & --4.8 &--0.4 & 64&63&47&--20.5\\
            \noalign{\smallskip} \hline
        \end{tabular}
        \begin{tablenotes}
            \item{$^\ast$} Ages for NGC~6913 and IC~4996 are from \citet{2023MNRAS.525.2315A}. For IC\,4996, the input parameters
            $r$ and $\mu$ are taken from the catalog of \citet{2024A&A...686A..42H}.

        \end{tablenotes}
    \end{threeparttable}
\end{table*}

To obtain the peculiar velocities $(U_\mathrm{s},V_\mathrm{s},W_\mathrm{s})$ with respect to the regional standard
of rest (RSR) centered at the positions of the clusters, we first calculated the velocities $(U,V,W)$ with respect
to Local Standard of Rest (LSR) centered at the Sun and then transformed them into a Cartesian Galactocentric
reference frame rotating with the Galaxy. The details of calculations and a full set kinematic parameters with
their errors are given in Appendix\,\ref{sec:appD} and its Table\,\ref{table:orbit_params}. In Table \ref{uvw} we
list only the input data and summarize those kinematic parameters which will be discussed in this and the next
subsection. For NGC\,6913, the values of mean distance and proper motion components are based on probable member
stars identified by us, and for IC\,4996 these two input parameters are taken from \citet{2024A&A...686A..42H}.
The internal velocity dispersions are not given in the table, since they do not differ significantly from cluster
to cluster: the observed dispersion in total proper motion is in the range 0.13\,--\,0.17 mas~yr$^{-1}$, or
0.8--1.1 {\kms} in 2D velocity.

As is seen in Table\,\ref{uvw}, the components of peculiar motion of the clusters are within $\sim$10 {\kms},
except for the component $U_\mathrm{s}$ of Berkeley 87. Positive values of $U_\mathrm{s}$, and the largest of the
three components, indicate a significant motion of all these clusters toward the Galactic center, especially
pronounced for Berkeley 87. The negative values of azimuthal component $V_\mathrm{s}$ indicate a moderate (up to
$\sim$10 {\kms}) motion counter to Galactic rotation. The vertical velocity $W_\mathrm{s}$ is most noticeable for
Berkeley 86 which appears to be heading upwards from the Galactic plane twice as fast as Berkeley 87 and
NGC\,6913.

In general, the kinematics of all four neighboring clusters are similar in that they are moving radially toward
the Galactic center and opposite to the Galactic rotation. Such kinematic behavior is observed for high-mass star
forming regions in the Local arm \citep[e.g.,][]{2013ApJ...769...15X} and other Galactic spiral arms
\citep{2014ApJ...790...99C,2019ApJ...885..131R,2019ApJ...874...94W}. This is consonant with the general concept in
spiral density wave theory whereby the formation of stars occurs from gas shocked when entering a spiral arm, with
newborn stars moving inward toward the Galactic center \citep[e.g.,][]{1969ApJ...158..123R}.

\subsection{Traceback in time} \label{subsec:traceback}

We traced back in time the trajectories $(X,Y,Z)$ and motions $(U,V,W)$ of the clusters from Table\,\ref{uvw}
using the \texttt{Galpy} package by \citet{2015ApJS..216...29B} with the Galaxy's potential model
\texttt{MWPotential2014} which consists of a spherical bulge with a power-law density profile, a Miyamoto-Nagai
disk, and a dark-matter halo described by a Navarro-Frank-White potential. Following the results for the Local Arm
and Galactic disk structure from \citet{2019ApJ...885..131R}, we adopted a Solar distance to the Galactic Center
of  $R_0 = 8.15$ kpc and a local circular velocity of $V_c(R_0) = 236.0$ km s$^{-1}$. For vertical positioning, we
adopted a Solar height of $z_\odot = 20.8$ pc above the Galactic midplane \citep{2016ARA&A..54..529B}, ensuring
alignment with the center of the Galactocentric coordinate system. The uncertainties in the positions with time
were estimated using a Monte Carlo (MC) technique with 10\,000 artificial orbits generated by adding a randomly
dispersed propagation of observational errors in $r$, $\mu_\alpha$, $\mu_\delta$, and RV.

The table with the derived orbital parameters in its entirety is given in Appendix\,\ref{sec:appD}. The distances
$(Z)$ of the clusters at present time ($t$=\,0) and at presumable time of their formation ($t=-t_{age}$), the
maximum height from the Galactic plane to be reached on their Galactic orbits, and the time of crossing the
Galactic mid-plane, $t_{Z=0}$, are also summarized in Table\,\ref{uvw}.

Both Berkeley clusters, along with NGC~6913, follow similar Galactic orbits, whereas the orbital motion of IC~4996
is quite different. This cluster, being more distant and likely of greater age, is separated from the rest three
OCs too widely ($>$\,200~pc), both at present time and throughout its lifetime; thus it cannot be considered as
part of the same kinematic subgroup. Berkeley 87 and NGC~6913 trace very closely their paths, with nearly the same
maximum height from the Galactic plane, $\sim$40~pc, to be reached on their orbits. Berkeley 86, however, can
reach $Z_{\rm max}$ twice that, $\sim$80 pc. Nevertheless, this cluster, like Berkeley 87 and NGC~6913, might have
been formed sufficiently close to the Galaxy's plane of symmetry ($Z_{\mathrm{birth}}$$\sim$15 pc). On the other
hand, the time of clusters crossing this plane, at $\sim$\,9--11 Myr from now, is not radically different from
their age estimates.

\begin{figure}
   \centering
   \includegraphics[width=\hsize,trim={2.5cm 1.1cm 3.0cm 2.2cm},clip]{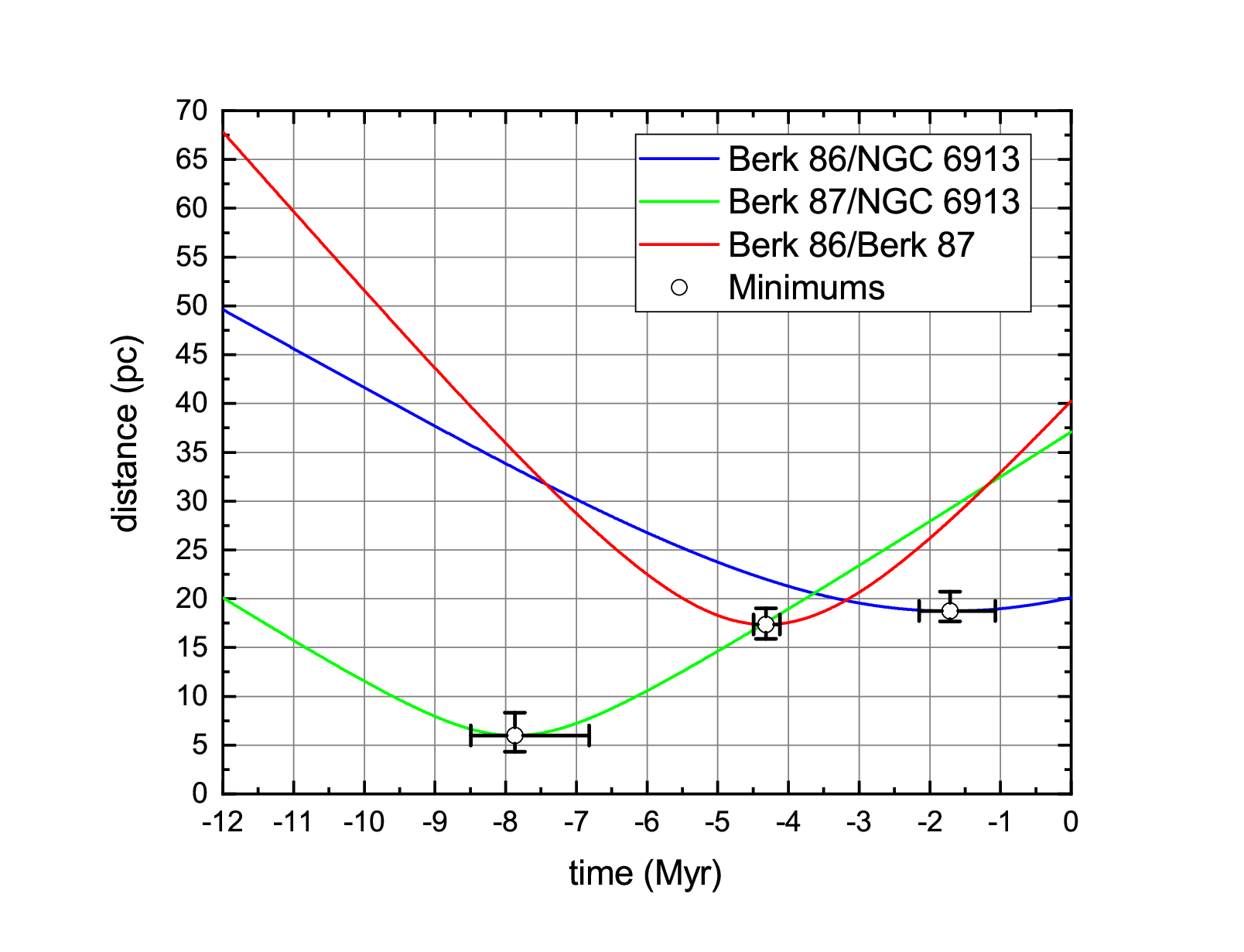}
      \caption{Traceback in time of the distance between the centers of the clusters  Berkeley\,86, Berkeey\,87, and NGC~6913. The error bars at the points of closest approach
      are 16\% and 84\% quantiles which delimit the range of uncertainties due to propagation of observational errors.}
      \label{fig:Myr}
   \end{figure}

With the evident similarity of the spatial positions, kinematic behavior, and ages, there is every reason to
expect that Berkeley~86, Berkeley~87, and NGC~6913 may have a common origin out of the same molecular cloud. To
see whether these clusters could have been formed close to one another, we plotted in Fig.\,\ref{fig:Myr} the
separations between their centers as a function of time backwards to 12 Myr. The time and the values of
separations at closest approach to each other are summarized in Table\,\ref{separ}; the errors given are from MC
simulations. It should be noted that the uncertainties in the adopted distances to the clusters and in the values
of RV contribute a major amount to the total uncertainty in the calculated values of closest separation and its
dating, while relatively small errors in proper motions have minimal impact.


  \begin{table}
      \caption[]{Center-to-center separations between the pairs of open clusters.
       $d_0$ is the present-day separation, and $d_{\rm min}$ is the separation at
                  closest approach at time $t_{d_{\rm min}}$. Uncertainties are from
                  16\% and 84\% quantiles from MC simulations.} \label{separ}
\centering
    \begin{tabular}{c r l r}
           \hline\hline \noalign{\smallskip}
\multicolumn{1}{c}{Pair of clusters} & \multicolumn{1}{c}{$d_0$} & \multicolumn{1}{c}{$d_{\rm min}$} &
\multicolumn{1}{c}{$t_{d_{\rm min}}$}\\
 & \multicolumn{1}{c}{(pc)} & \multicolumn{1}{c}{(pc)} & \multicolumn{1}{c}{(Myr)}\\
    \hline
    \noalign{\smallskip} \noalign{\smallskip}
  Berkeley~86\,/Berkeley~87 & 40.7$^{+4.5}_{-1.5}$ & 17.3$^{+1.7}_{-1.5}$ & $-4.3^{+0.2}_{-0.2}$\\
            \noalign{\smallskip} \noalign{\smallskip}
  Berkeley~86\,/\,NGC~6913 & 21.4$^{+4.5}_{-1.4}$ & 18.7$^{+2.0}_{-1.1}$ & $-1.7^{+0.6}_{-0.4}$\\
            \noalign{\smallskip} \noalign{\smallskip}
  Berkeley~87\,/\,NGC~6913 & 37.2$^{+5.6}_{-2.2}$ & ~~6.0$^{+2.3}_{-1.7}$ & $-7.9^{+1.1}_{-0.6}$\\
\noalign{\smallskip} \hline
         \end{tabular}
   \end{table}

Consider first the pair Berkeley~87 and NGC\,6913, which reaches the minimum center-to-center distance,
$\sim$\,6$\pm$2 pc, at $\approx$\,8$\pm$1 Myr from now, i.e., at time close, within the errors, to the age dated
for these clusters. A very close approach, comparable with the present size of each of the two clusters
\citep[their radii are $\sim$\,5 pc, see][]{2024A&A...686A..42H}, similar ages, and common kinematics, all tied
together, strongly suggest that they might have been born in pair from primordial fragmentation of a common gas
cloud.

With the adopted values of line-of-sight distances to Berkeley 86 and 87, tracing their orbits back in time yields
the closest center-to-center separation at $\sim$\,4.3$\pm$0.2 Myr from now, i.e. in the more recent past than
their age expected from isochrone data ($\sim$\,6--8 Myr). On the other hand, the orbital path of NGC~6319
relative to those of Berkeley 86 and 87 shows no indication that these two Berkeley clusters might be born in the
same place.

Consider now the time at $\sim$\,4 Myr from now, at which the orbital paths of all three clusters enter in
approach of $\sim$20 pc (see Fig.\,\ref{fig:Myr}). Such separation is just below the combined tidal radius of the
clusters, if to take into account their total observed masses given by \citet{2023MNRAS.525.2315A} or even smaller
masses obtained by \citet{2024A&A...686A..42H}. Therefore we cannot rule out the possibility that around that time
in the past all these three clusters might be weakly interacting. Had that been the case, the less populated and
less massive Berkeley 86 would perhaps have met shredding of part of its stars. Dynamical considerations are
beyond the scope of the present paper, but we believe that the dynamical evolution of these clusters would be an
important topic for future investigations.

\section{Relation to the Cygnus OB1 association} \label{subsec:CygOB1}

The young open clusters Berkeley 86 and 87, along with their neighbors NGC\,6913 and IC\,4996, have long been referred to
as part of the Cygnus OB1 association \citep[e.g.,][ among many others]{1978ApJS...38..309H,1992A&AS...94..211G,2015MNRAS.454.2486S}.
However \citet{2021MNRAS.508.2370Q,2022MNRAS.511.1224Q} in their study based on {\it Gaia} DR2 data argue that most of the known OB
associations in Cygnus, including Cyg~OB1, do not show the kinematic coherence expected for OB associations. Instead,
these authors identified kinematically six new association-like groups in Cygnus. Therefore, we consider below the
relationship of the two Berkeley clusters to Cyg~OB1 as well as to new associations identified in the above-mentioned
study of Quintana \& Wright.

\begin{figure}
    \centering
    \includegraphics[width=\hsize,trim={0.5cm 1.1cm 1.2cm 2.3cm},clip]{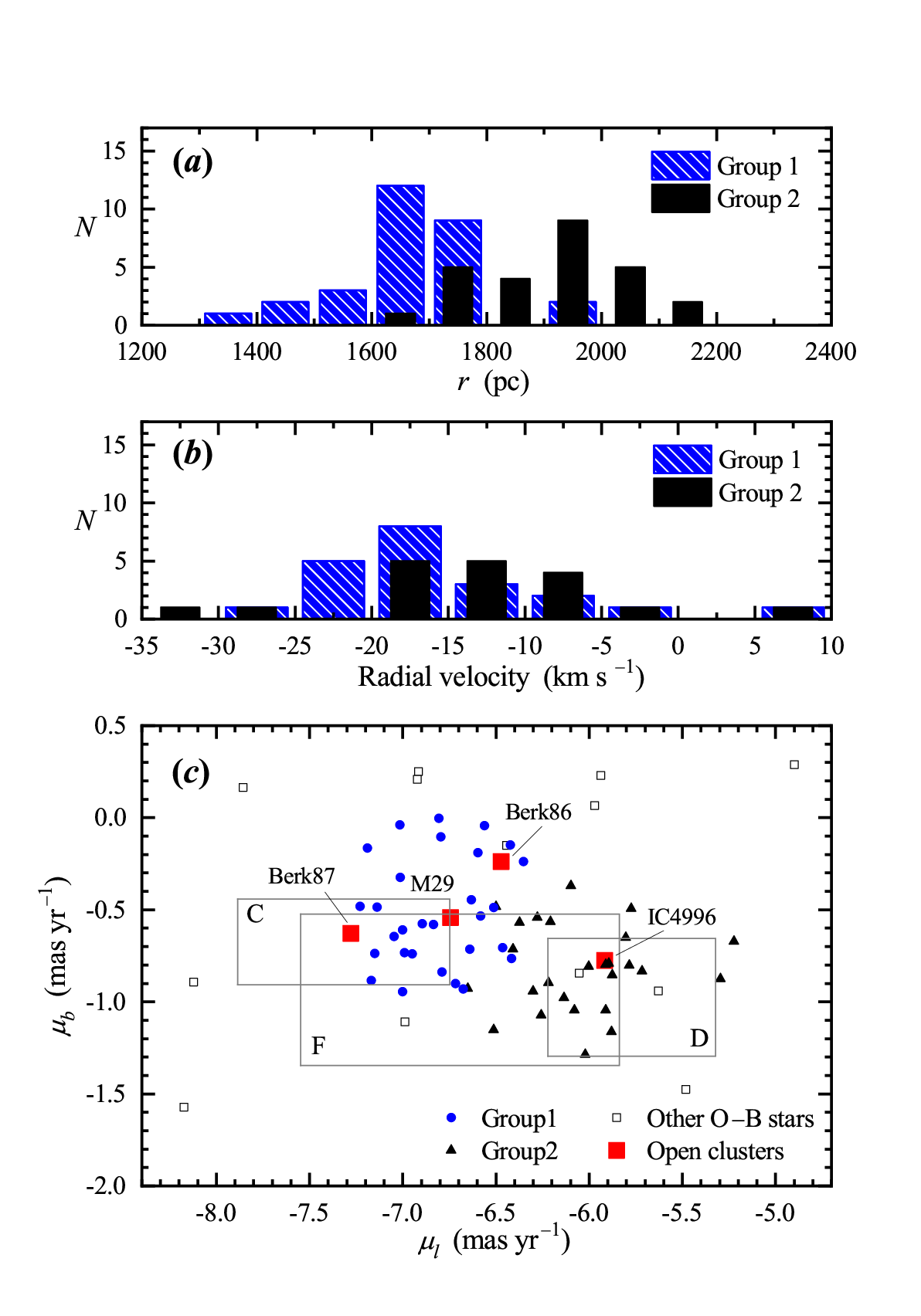}
    \caption{Stars of the Cyg OB1 association: (\emph{a}) the distribution by distance $rpgeo$ in the range
        1.3--2.5 kpc; (\emph{b}) the distribution by RV; (\emph{c}) the vector point diagram of proper motions. In panel (\emph{c}), small open squares denote O--B stars not used
        to estimate the association properties given in Table\,\ref{OB1tab}, three large rectangles mark the boundaries of
        overlapping associations (C, D, F) identified by \citet{2021MNRAS.508.2370Q} from {\it Gaia} DR2 data.
    } \label{fig:OB1rpm}
\end{figure}

As a major source of Cyg~OB1 stars, the list from \citet{1989AJ.....98.1598B} was used, which we supplemented with
Strasbourg CDS data (accurate coordinates and MK spectral types) and then cross-matched with the {\it Gaia} DR3
catalog. After eliminating from the list the stars that have been identified as probable members of OCs and adding
new stars with zero cluster membership probabilities (in both cases, based on {\it Gaia} DR3 data), we compiled a
list of 73 O--B stars in the distance range from 1.3 to 2.5 kpc. The majority of these stars have published RV
data; we have only gathered them from the literature and averaged multiple RV measurements on a given star,
rejecting those with large dispersion or large measured uncertainties. The distributions of the association stars
by distance $rpgeo$ and RV, and their vector point diagram (VPD) are plotted in Fig.\,\ref{fig:OB1rpm}.

\begin{figure*}
    \centering
    \includegraphics[width=\hsize,trim={1.0cm 1.25cm 2.5cm 7.0cm},clip]{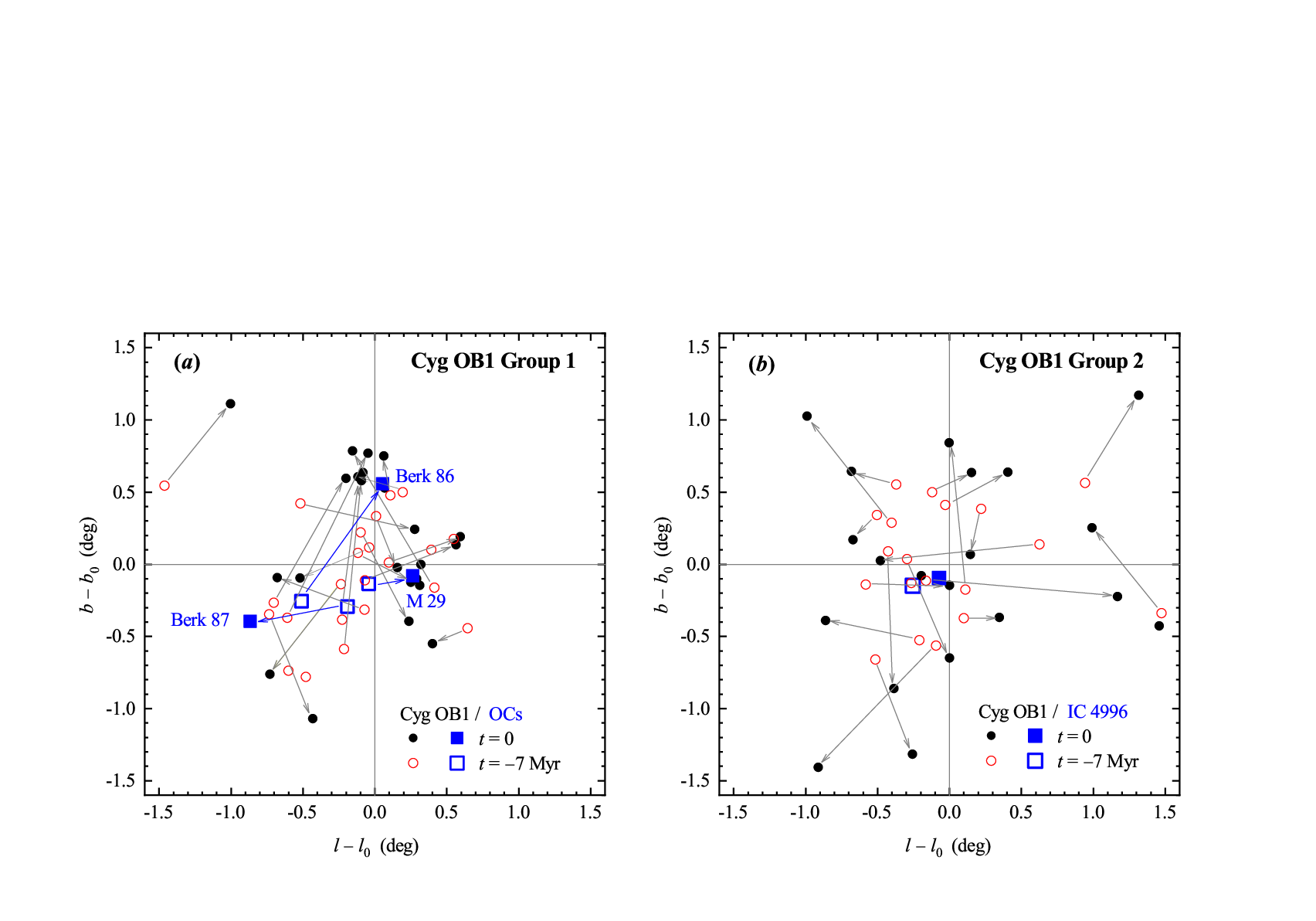}
    \caption{ Positions of stars and open clusters in two groups of the Cyg~OB1 association at present time
        (filled symbols) and 7 Myr ago (open symbols). The center of each group is at the origin of galactic coordinates
        which are defined as they would be measured by an observer located at the position of the Sun and moving with the
        circular velocity along the solar circle. The squares represent the OCs associated with a group: NGC\,6913\,(M\,29), Berkeley 86, and
        Berkeley 87 in panel (\emph{a}), and IC\,4996 in panel (\emph{b}). The arrows illustrate the shifts in positions
        from $t$\,=\,--7 Myr to present time.} \label{fig:OB1arr}
\end{figure*}

\begin{table*}
    \begin{threeparttable}
        \caption[]{Properties of two groups of Cygnus OB1 from present paper and overlapping associations C and D from \citet{2021MNRAS.508.2370Q}.
        The subscript `m' indicates the median value, $\sigma_{v_\ell}$ and $\sigma_{v_b}$ denote the
                tangential velocity dispersions. $N_{\rm stars}$ is the number of stars used in the calculation of median values,
                except for the RV column headed ``$v_{r_{\rm m}}$'' where the number of stars used is given in parentheses following the
                RV value. The values cited for associations C and D, based on {\it Gaia} DR2, are given in italics.} \label{OB1tab}
        \centering{\footnotesize                         
            \begin{tabular}{l c c r r r r r r c c c}
                \hline\hline \noalign{\smallskip}
                \multicolumn{1}{l}{Substructure} & \multicolumn{1}{r}{RA(2000)$_{\rm m}$} & \multicolumn{1}{c}{Dec(2000)$_{\rm m}$} &
                \multicolumn{1}{c}{$\ell_{\rm m}$} & \multicolumn{1}{c}{$b_{\rm m}$} & \multicolumn{1}{c}{$d_{\rm m}$} &
                \multicolumn{1}{c}{$\mu_{\ell_{\rm m}}$} & \multicolumn{1}{c}{$\mu_{b_{\rm m}}$} & \multicolumn{1}{c}{$v_{r_{\rm m}}$} &
                \multicolumn{1}{c}{~~~$\sigma_{v_\ell}$} & \multicolumn{1}{c}{$\sigma_{v_b}$} & \multicolumn{1}{c}{$N_{\rm stars}$}\\
                \multicolumn{1}{l}{of Cygnus OB1}     & \multicolumn{1}{c}{(deg)} & \multicolumn{1}{c}{(deg)} & \multicolumn{1}{c}{(deg)} &
                \multicolumn{1}{c}{(deg)} & \multicolumn{1}{c}{(pc)} &  \multicolumn{1}{c}{(mas)} & \multicolumn{1}{c}{(mas)} &
                \multicolumn{1}{c}{(\kms)} & \multicolumn{1}{c}{~~~(\kms)} & \multicolumn{1}{c}{~~~(\kms)} & \\
                \hline \noalign{\smallskip} \noalign{\smallskip}
                Group1 & 305.57 & 38.52 & 76.61 & 0.73 & 1688 & --6.81 & --0.58 & --17.0$\pm$2.3~(2\rlap{1)} & 2.2 & 2.3 & 29\\
                \it{Association C}  & \it{305.45} & \it{37.77} & \it{76.02} & \it{0.54} & \it{1710} & \it{--7.01} & \it{--0.61} & \it{--19.2$\pm$3.0~(1\rlap{0)}} & \it{2.2} & \it{0.9} & \it{69}\\
                \noalign{\smallskip}
                Group2 & 304.29& 37.76 & 75.45 & 1.40 & 1930 & --6.05 & --0.82 & --12.7$\pm$2.7~(1\rlap{8)} & 3.1& 2.1 & 26\\
                \it{Association D}  & \it{304.28} & \it{37.65} & \it{75.42} & \it{1.24} & \it{1999} & \it{--5.83} & \it{--0.87} & \it{--7.0$\pm$3.9~(1\rlap{0)}} & \it{1.5} & \it{1.4} & \it{60}\\
                \noalign{\smallskip} \hline
            \end{tabular}
        }
        \begin{tablenotes}
            \item{}
        \end{tablenotes}
    \end{threeparttable}
\end{table*}

Examination of the three panels in Fig.\,\ref{fig:OB1rpm} shows that both Berkeley clusters and NGC\,6913 are
likely to be part of the same kinematic substructure of this association, while the cluster IC\,4996 represents
the core of another expanding substructure. Here we denote these substructures as Group 1 and Group 2. In earlier
analysis of the kinematical structure of Cyg~OB1 in the phase subspace of angular coordinates and RVs
\citet{2017MNRAS.465.3879C} also considered IC~4996 as representative of separate stellar group within Cyg~OB1,
distinct from the other segregated group related to the cluster Dolidze~42 and its neighbors NGC\,6913, Berkeley
86 and Berkeley 87. Three new association-like groups (C, D, F) identified by
\citet{2021MNRAS.508.2370Q,2022MNRAS.511.1224Q} cover the main domains of known Cyg~OB1 stars in the VPD in
Fig.\,\ref{fig:OB1rpm}(\emph{c}). However, very few stars are found in common with the members of their groups C
(four stars), D (two stars), and F (one star). The properties of the two groups of Cygnus OB1 stars are summarized
in Table\,\ref{OB1tab}, where we also compare them with the properties of overlapping associations C and D. It is
evident from the table, that both substructures of O--B stars can be associated with, or are part of, the
associations C and D, as most of the median values of properties appear to be similar.

As an illustration of expansion of the Cygnus OB1 stars and associated clusters, we plotted in
Fig.\,\ref{fig:OB1arr} their positions projected on the sky at present time and at --7 Myr from now. Panel
(\emph{a}) shows that Group 1 with the two Berkeley clusters and NGC\,6913 (panel \emph{a}) represent a relatively
compact grouping of O--B stars with a diameter of about 50 pc. It is interesting to note the absence of O--B stars
at present time around the massive cluster Berkeley 87, whereas most candidates of this group are cospatial with
the rest two OCs, Berkeley 86 and NGC\,6913. This may indicate that part of O--B stars in this group were original
members of the latter two clusters, that were eventually lost from their parental clusters to form halos around
them or to become members of unbound association. The other substructure of Cyg OB1, Group 2 with IC\,4996 in its
core (panel \emph{b}), is more impressive in that it clearly exhibits a kinematic pattern of expansion: the
observed diameter of this substructure is about 100 pc, while that traced back to --7 Myr becomes of only $\sim$45
pc.

\section{Conclusions}  \label{sec:concl}
In this work, we investigate the properties of two young open clusters in Cygnus, Berkeley 86 and Berkeley 87, by
using the {\it Vilnius} seven-color photometry down to $V$\,=19 mag, combined with the {\it Gaia} DR3 astrometry.
We identified 63 and 92 probable members ($p\geq 0.70$) in each of the clusters, respectively.

Both clusters are almost equidistant (1691 and 1681 pc). The mean interstellar extinction toward their member
stars is $A_V$=2.57$\pm\,$0.38 and 4.09$\pm\,$0.39 mag, respectively. Beyond the clusters, the extinction in $V$
can reach $\approx$\,9.1 mag at 4--5 kpc, as evaluated using, in addition, the 2MASS and WISE photometry of red
clump giants.

The extinction is nonuniform across the face of each of the clusters, with the differences from star to star
amounting at most to 1.5 mag in $A_V$. In Berkeley 86, the more-heavily reddened stars all lie on the south side
of the cluster, where the high-mass stars and stars of possibly younger ages are located. This suggests that the
southern area might have been the site of later stage of star formation in this cluster. In Berkeley 87, the
more-heavily reddened stars, many of which can also be younger than the rest members, are predominantly located
around the center and to the east of it, i.e. nearer to the nearby dust complex.

The total mass of Berkeley 86 is estimated to be 519 $\mathrm{M}_\sun$, and that of Berkeley 87, 1551
$\mathrm{M}_\sun$. The population of both clusters is dominated by pre-main-sequence stars, with their average age
6.1$\pm0.5$ Myr and a dispersion of $\sim$3 Myr. For the most massive stars ($\geq$11 $\mathrm{M}_\sun$) in
Berkeley 87 we obtain the age 8.9$\pm1.3$ Myr. The latter cluster can, on average, be slightly older than Berkeley
86.


The kinematics both Berkeley clusters is closely similar to that of other two neighboring young clusters in the
region, NGC\,6913 and IC\,4996, as they all are moving radially toward the Galactic center and counter to the
Galactic rotation. A comparison with the kinematics of the known members of the Cyg~OB1 association indicates that
the former three clusters are likely to be part of the same kinematic substructure of this association, while
IC\,4996 represents the core of another expanding substructure of Cyg~OB1.

The orbital paths show that at $\sim$4 Myr from now the two Berkeley clusters and NGC\,6913 might have entered in
approach to each other, what could lead to weak interaction and shredding of some stars of the less massive
Berkeley 86. It seems unlikely that both Berkeley clusters could have had a common birthplace, however Berkeley~87
and NGC\,6913 might have been born in pair. A study of chemical composition would be a valuable further step to
test whether the latter two clusters do indeed represent a genetic pair. \vskip 5mm

\begin{acknowledgements}
    It is a pleasure to thank Stanislava Barta\v{s}i\={u}t\.{e} for valuable remarks and guidance regarding kinematics.
    This research has made use of the SIMBAD database, operated at CDS, Strasbourg, France
    \citep{2000A&AS..143....9W}.
    This work has made use of data from the European Space Agency (ESA) mission
    {\it Gaia} (\url{https://www.cosmos.esa.int/gaia}), processed by the {\it Gaia}
    Data Processing and Analysis Consortium (DPAC,
    \url{https://www.cosmos.esa.int/web/gaia/dpac/consortium}). Funding for the DPAC
    has been provided by national institutions, in particular the institutions
    participating in the {\it Gaia} Multilateral Agreement.
    This research has made use of the Astrophysics Data System, funded by NASA under Cooperative Agreement 80NSSC21M0056.
    Part of this work was supported by funding from the Research Council of Lithuania (LMTLT, grant No. S-MIP-23-89).

\end{acknowledgements}

\bibliographystyle{aa} 
\bibliography{b86b87}


\begin{appendix}

\onecolumn
\section{Tables of CCD photometry, classification and cluster membership}
 \label{sec:appX}

\begin{table}[ht]
    \setlength{\tabcolsep}{3.0pt}
    \caption{The catalog of {\it Vilnius} CCD photometry and derived photometric spectral types for 1053 stars in the Berkeley 86 area.}
    \label{table:1}
    \footnotesize
    \begin{tabular}{rccrcccccccccccccl}
        \hline\hline \noalign{\smallskip}
        {No.} &  {RA(J2000)} &  {DEC(J2000)} &  {$V$} &
        {$U$--$V$} &  {$P$--$V$} &  {$X$--$V$} &  {$Y$--$V$} &
        {$Z$--$V$} &  {$V$--$S$} &  {e$_V$} &  {e$_{(U-V)}$} &
        {e$_{(P-V)}$} &  {e$_{(X-V)}$} &  {e$_{(Y-V)}$} &
        {e$_{(Z-V)}$} &  {e$_{(V-S)}$} &{Photom.} \\
        { } &  {h~~~~m~~~~~s~} &  {$\circ$~~~~$\prime$~~~~$\prime\prime$} &
        {mag} &  {mag} &  {mag} &  {mag} &  {mag} &  {mag} &
        {mag} &  {mag} &  {mag} &  {mag} &  {mag} &  {mag} &
        {mag} &  {mag} &  {sp. type} \\
        \hline\noalign{\smallskip}

        1 & 20:19:50.43 & +38:38:08.8 & 16.385 &  ""   & 3.361 & 2.157 & 0.829 & 0.362 & 0.828 & 0.059 &  ""   & 0.161 & 0.091 & 0.077 & 0.072 & 0.075 & K0 V  \\
        2 & 20:19:50.84 & +38:46:54.7 & 16.423 & 3.546 & 2.820 & 1.988 & 1.072 & 0.348 & 0.889 & 0.016 & 0.032 & 0.031 & 0.028 & 0.038 & 0.032 & 0.037 & F4 V  \\
        3 & 20:19:51.02 & +38:46:21.3 & 17.169 & 3.956 & 3.316 & 2.3   & 1.073 & 0.381 & 1.01  & 0.016 & 0.038 & 0.033 & 0.029 & 0.042 & 0.034 & 0.041 & G4 V  \\
        ... &  &  &  &  &  &   &  &  &   &  &  &  &  &  &  &  &   \\
        \hline
    \end{tabular}
    \tablefoot{Table E.1 is published in its entirety in the machine-readable format.
        A portion is shown here for guidance regarding its form and content.}
\end{table}
\begin{table}[ht]
    \setlength{\tabcolsep}{3.0pt}
    \caption{The catalog of {\it Vilnius} CCD photometry and derived photometric spectral types for 502 stars in the Berkeley 87 area.}
    \label{table:2}
    \footnotesize
    \begin{tabular}{rccrcccccccccccccl}
        \hline\hline \noalign{\smallskip}
        {No.} &  {RA(J2000)} &  {DEC(J2000)} &  {$V$} &
        {$U$--$V$} &  {$P$--$V$} &  {$X$--$V$} &  {$Y$--$V$} &
        {$Z$--$V$} &  {$V$--$S$} &  {e$_V$} &  {e$_{(U-V)}$} &
        {e$_{(P-V)}$} &  {e$_{(X-V)}$} & \ {e$_{(Y-V)}$} &
        {e$_{(Z-V)}$} &  {e$_{(V-S)}$} &{Photom.} \\
        { } &  {h~~~~m~~~~~s~} & {$\circ$~~~~$\prime$~~~~$ \prime\prime$} &
        {mag} &  {mag} &  {mag} &  {mag} &  {mag} &  {mag} &
        {mag} &  {mag} &  {mag} &  {mag} &  {mag} &  {mag} &
        {mag} &  {mag} &  {sp. type} \\
        \hline\noalign{\smallskip}
        1 & 20:21:09.88 & +37:25:20.2 & 16.354 &  ""   &  ""   & 4.179 & 1.889 & 0.736 & 1.659 & 0.016 &  ""    & ""    & 0.031 & 0.077 & 0.030 & 0.025 & K1 III \\
        2 & 20:21:10.66 & +37:29:11.7 & 15.694 &  ""   &  ""   & 2.242 & 1.063 & 0.357 & 0.992 & 0.021 &  ""    & ""    & 0.024 & 0.059 & 0.027 & 0.026 & G1 V   \\
        3 & 20:21:10.75 & +37:21:23.6 & 17.915 & 5.208 & 4.193 & 3.082 & 1.565 & 0.574 & 1.381 & 0.014 &  0.067 & 0.039 & 0.039 & 0.018 & 0.020 & 0.021 & F8 IV  \\
        ... &  &  &  &  &  &   &  &  &   &  &  &  &  &  &  &  &   \\
        \hline
    \end{tabular}
    \tablefoot{Table E.2 is published in its entirety in the machine-readable format.
        A portion is shown here for guidance regarding its form and content.}
\end{table}

\begin{table}[ht]
    \centering \setlength{\tabcolsep}{3.5pt} \caption{The list of 63 probable members of Berkeley 86. }
    \label{table:3}
    \begin{tabular}{ccccccccccclcc}
        \hline\hline \noalign{\smallskip}
            {No.} &  {$Gaia$ $DR3$} &  {$\mu_{\alpha}$} &
            {$e_{\mu\alpha}$} &  {$\mu_{\delta}$} &  {$e_{\mu\delta}$} &
            {$rpgeo$} &  {$b_{rpgeo}$} &  {$B_{rpgeo}$} &  {$p$} &
            {$p_{\mathrm{HR}}$} &  {Sp. type} &  {$A_V$} &  {$\sigma_Q$} \\
            { } &  {source ID} &  {mas/yr} &  {mas/yr} &
            {mas/yr} &  {mas/yr} &  {pc} &  {pc} &  {pc} &
            { } &  { } &  { } &  {mag} &  {mag} \\
            \hline\noalign{\smallskip}
        19   & 2061290358837958656 & -3.525 & 0.024 & -5.495 & 0.030 & 1812 & 1748 & 1894 & 0.97 & "" & F3 V   & 1.98 & 0.10 \\
        23   & 2061290358837364864 & -3.609 & 0.030 & -5.349 & 0.038 & 1616 & 1543 & 1700 & 0.99 & "" & G7 IV  & 2.82 & 0.07 \\
        48   & 2061292351702257536 & -3.476 & 0.019 & -5.404 & 0.024 & 1703 & 1652 & 1762 & 0.99 & 0.99 & A1 V   & 2.21  & 0.03 \\
        ...&  &  &  &  &  &  &  &  &  &  &  &  &  \\
        \hline
    \end{tabular}
    \begin{tablenotes}
        \item{}{\footnotesize {\bf Description of Columns:}
        \item{\textit{Col. (1)}}~~Star numbers as in the catalog of photometry (Table \ref{table:1}).
        \item{\textit{Col. (2)}}~~Source ID.
        \item{\textit{Cols. (3)--(6)}}~~Proper motion components and their errors from {\it Gaia} DR3.
        \item{\textit{Cols. (7)--(9)}}~~Distances $rpgeo$ from \citet{2021AJ....161..147B}.
        \item{\textit{Col. (10)}}~~Membership probabilities from this work ($p$).
        \item{\textit{Col. (11)}}~~Membership probabilities from \citet{2024A&A...686A..42H} ($p_{\mathrm{HR}}$).
        \item{\textit{Col. (12)}}~~Spectral type from {\it Vilnius} photometry.
        \item{\textit{Col. (13)}}~~Extinction $A_V$ from {\it Vilnius} photometry.
        \item{\textit{Col. (14)}}~~Quality of photometric classification.}
    \end{tablenotes}
    \tablefoot{
        Table E.3 is published in its entirety in the machine-readable format.
        A portion is shown here for guidance regarding its form and content.}
\end{table}

\begin{table}[ht]
    \centering \setlength{\tabcolsep}{3.5pt} \caption{The list of 92 probable members of Berkeley 87. Star numbers as
        in the catalog of photometry (Table \ref{table:2}). Description of other columns as in Table\,\ref{table:3}.}
    \label{table:4}
    \begin{tabular}{ccccccccccclcc}
            \hline\hline \noalign{\smallskip}
            {No.} & {$Gaia$ $DR3$} &  {$\mu_{\alpha}$} &
            {$e_{\mu\alpha}$} &  {$\mu_{\delta}$} &  {$e_{\mu\delta}$} &
            {$rpgeo$} &  {$b_{rpgeo}$} &  {$B_{rpgeo}$} &  {$p$} &
            {$p_{\mathrm{HR}}$} &  {Sp. type} &  {$A_V$} &  {$\sigma_Q$} \\
            { } &  {source ID} &  {mas/yr} &  {mas/yr} &
            {mas/yr} &  {mas/yr} &  {pc} &  {pc} &  {pc} &
            { } &  { } &  { } &  {mag} &  {mag} \\
            \hline \noalign{\smallskip}
        4 & 2060968923471305856 & -3.606 & 0.018 & -6.353 & 0.018 & 1635 & 1581 & 1676 & 0.99 & 1.00 & A0 V   & 3.99 & 0.03 \\
        5 & 2060969713745296256 & -3.736 & 0.041 & -6.558 & 0.047 & 1599 & 1494 & 1734 & 0.97 & 1.00 & A4 V   & 4.36 & 0.05  \\
        22 &2057960075538765312 & -3.460 & 0.017 & -6.334 & 0.018 & 1516 & 1481 & 1556 & 0.96 & 0.7 & B9 V   & 2.58 & 0.03 \\
        ...&  &  &  &  &  &  &  &  &  &  &  &  &  \\
        \hline
    \end{tabular}
    \tablefoot{Table E.4 is published in its entirety in the machine-readable format. A portion is shown here for
        guidance regarding its form and content.}
\end{table}
\onecolumn

    \section{Literature data on the open clusters Berkeley 86 and Berkeley 87} \label{sec:appA}
    \begin{table}[h]
        \caption{Literature data on Berkeley 86. Bolded values are derived
            by the cited authors, and not bolded values are cited from previous papers.}
        \label{table:A1}
        \footnotesize
        \begin{tabular}{lccccccl}
            \hline\hline \noalign{\smallskip}
            {Bib source} &  {Number} &
            {distance} & {pmRA} & {pmDEC} & {Age} & {RadVel} & {E(B-V)$^\ast$} \\
        {(sorted by year)} & {of stars} & {(pc)} & {(mas/yr)} & {(mas/yr)} & {(Myr)} & {(km/s)} & {(mag)} \\
        \hline \noalign{\smallskip}\noalign{\smallskip}
        \citet{1981PASP...93..441F} & 8 & \textbf{1720} & ~ & ~ & \textbf{6} & ~ & \textbf{0.96$\pm$0.07} \\
        \noalign{\smallskip}
        \citet{1991MNRAS.249...76B} & 11 & 1720 & ~ & ~ & ~ & ~ & \textbf{0.99} \\
        \noalign{\smallskip}
        \citet{1992AJ....103..916F} & 22 & \textbf{1590} & ~ & ~ & \textbf{5$\pm$1} & -22$\pm$7 & \textbf{1.01 (0.8-1.2)} \\
        \noalign{\smallskip}
        \mbox{\citet{1993A&AS..101...37C}} & ~ & 1110 & ~ & ~ & 40 & ~ & ~ \\
        \noalign{\smallskip}
        \citet{1995ApJ...454..151M} & 10 & \textbf{1900} & ~ & ~ & 2-3 & ~ & \begin{tabular}{@{}c@{}}$\begin{cases}\textbf{0.80$\pm$0.04} \\ \textbf{(0.63-0.92)}\end{cases}$\end{tabular} \\
        \noalign{\smallskip}
        \mbox{\citet{1996A&AS..119..221D}} & 126 & ~ & ~ & ~ & ~ & ~ & \textbf{1.02 (0.9-1.25)} \\
        \noalign{\smallskip}
        \citet{1997AJ....113..713D} & 40 & \textbf{1660} & ~ & ~ & \textbf{8.7} & ~ & \textbf{0.89} \\
        \noalign{\smallskip}
        \citet{1997AstL...23...71G} & 7 & \textbf{790} & \textbf{-9.3$\pm$2.9} & \textbf{-11.1$\pm$1.9} & ~ & ~ & ~ \\
        \noalign{\smallskip}
        \citet{1999AstL...25..595R} & ~ & \textbf{780} & \textbf{-7.1} & \textbf{-11.7} & ~ & -19.3$\pm$7.8 & ~ \\
        \noalign{\smallskip}
        \mbox{\citet{1999A&A...349..825V}} & ~ & 1660 & ~ & ~ & 4--6 & ~ & \textbf{(0.96--1.04)} \\
        \noalign{\smallskip}
        \mbox{\citet{2000A&AS..146..251B}} & 2 & \textbf{580} & \textbf{-3.8$\pm$0.54} & \textbf{-4.11$\pm$0.57} & ~ & ~ & ~ \\
        \noalign{\smallskip}
        \citet{2001AJ....121.1050M} & ~ & 1900 & ~ & ~ & \textbf{3.4}& ~ & 0.8 (0.6-0.9) \\
        \noalign{\smallskip}
        \citet{2001MNRAS.328..370Y} & 64 & 1100 & ~ & ~ & 12.6 & ~ &\textbf{0.74 (0.24-1.1)} \\
        \noalign{\smallskip}
        \mbox{\citet{2002A&A...392..869L}} & 67 & 1660 & ~ & ~ & ~ & ~ & \textbf{(0.8-1.1)} \\
        \noalign{\smallskip}
        \citet{2002NewA....7..553T} & 220 & \textbf{1161} & ~ & ~ & \textbf{10} & ~ & \textbf{0.7} \\
        \noalign{\smallskip}
        \mbox{\citet{2002A&A...389..871D}} & 37 & \textbf{1112} & \textbf{-4.56$\pm$0.01} & \textbf{-5.56$\pm$0.01} & \textbf{13} & \textbf{-25.54} & \textbf{0.898} \\
        \noalign{\smallskip}
        \mbox{\citet{2002A&A...390..945K}} & 11 & \textbf{1660} & ~ & ~ & \textbf{3--5} & ~ & ~ \\
        \noalign{\smallskip}
        \citet{2003AJ....125.1397C} & ~ & ~ & -3.8 & -4.11 & \textbf{10} & -19.3 & ~ \\
        \noalign{\smallskip}
        \citet{2003ARep...47....6L} & 25 & 1112 & \textbf{-3.06$\pm$0.32} & \textbf{-4.15$\pm$0.18} & 13 & ~ & ~ \\
        \noalign{\smallskip}
        \mbox{\citet{2005A&A...438.1163K}} & 4 & 1112 & \textbf{-6.17$\pm$0.91} & \textbf{-3.11$\pm$0.93} & \textbf{9.1} & -13.6$\pm$8.2 & 0.9 \\
        \noalign{\smallskip}
        \mbox{\citet{2006A&A...446..949D}} & 41 & ~ & \textbf{-4.15$\pm$0.44} & \textbf{-5.76$\pm$0.44} & ~ & ~ & ~ \\
        \noalign{\smallskip}
        \citet{2007BASI...35..383B} & 191 & \textbf{1585} & ~ & ~ & \textbf{6} & ~ & \textbf{0.95} \\
        \noalign{\smallskip}
        \citet{2008AJ....136..118F} & 2 & 1112 & \textbf{-3.8$\pm$1.3} & \textbf{-4.6$\pm$1.2} & 13.1 & \textbf{-25.54$\pm$2.64} & ~ \\
        \noalign{\smallskip}
        \citet{2009MNRAS.399.2146W} & ~ & 1112 & \textbf{-4.09$\pm$0.32} & \textbf{-4.54$\pm$0.32} & 13.1 & -25.5$\pm$2.6 & ~ \\
        \noalign{\smallskip}
        \mbox{\citet{2012A&A...541A..95F}} & 16 & 1585 & ~ & ~ & ~ & ~ & \begin{tabular}{@{}c@{}}$\begin{cases}\textbf{0.98; 0.89} \\ \textbf{(0.55--0.81)}\end{cases}$\end{tabular} \\
        \noalign{\smallskip}
        \mbox{\citet{2013A&A...558A..53K}} & ~ & \textbf{1653} & \textbf{-4.4} & \textbf{-2.17} & \textbf{6} & -19.3$\pm$7.8 & \textbf{0.958} \\
        \noalign{\smallskip}
        \citet{2017MNRAS.470.3937S} & 60 & 1112 & \textbf{-3.51} & \textbf{-5.84} & \textbf{13.2} & ~ & 0.9 \\
        \noalign{\smallskip}
        \citet{2018MNRAS.478.5184D} & 80 & ~ & \textbf{-3.12$\pm$1.1} & \textbf{-4.55$\pm$0.99} & ~ & ~ & ~ \\
        \noalign{\smallskip}
        \mbox{\citet{2018A&A...618A..93C}} & 31 & \textbf{1703} & \textbf{-3.447$\pm$0.066} & \textbf{-5.402$\pm$0.090} & ~ & ~ & ~ \\
        \noalign{\smallskip}
        \mbox{\citet{2019A&A...623A..80C}} & 1 & ~ & ~ & ~ & ~ & \textbf{151.9$\pm$0.83} & ~ \\
        \noalign{\smallskip}
        \citet{2019MNRAS.486.5726D} & 143 & \textbf{1750} & \textbf{-3.51$\pm$0.022} & \textbf{-5.497$\pm$0.027} & ~ & \textbf{-3.733$\pm$0.671} & ~ \\
        \noalign{\smallskip}
        \citet{2019ApJS..245...32L} & 72 & \textbf{1776} & \textbf{-3.468$\pm$0.229} & \textbf{-5.507$\pm$0.285} & \textbf{5$\pm$0.3} & ~ & ~ \\
        \noalign{\smallskip}
        \citet{2020MNRAS.495.1209R} & 11 & \textbf{1760} & ~ & ~ & ~ & ~ & ~ \\
        \noalign{\smallskip}
        \mbox{\citet{2020A&A...633A..99C}} & 31 & 1703 & -3.447$\pm$0.066 & -5.402$\pm$0.09 & ~ & ~ & ~ \\
        \noalign{\smallskip}
        \mbox{\citet{2020A&A...640A.127Z}} & ~ & 1703 & -3.447 & -5.402 & 6.8 & 12.25$\pm$22.23 & 0.958 \\
        \noalign{\smallskip}
        \mbox{\citet{2020A&A...640A...1C}} & 23 & 1621 & -3.447$\pm$0.066 & -5.402$\pm$0.09 & 11 & ~ & Av: 2.77 \\
        \noalign{\smallskip}
        \citet{2021ApJ...923..129J} & 130 & \textbf{1702} & \textbf{-3.456$\pm$0.055} & \textbf{-5.454$\pm$0.141} & ~ & ~ & ~\\
        \noalign{\smallskip}
        \mbox{\citet{2021A&A...651A.104P}} & 23 & \textbf{1621} & \textbf{-3.482$\pm$0.067} & \textbf{-5.416$\pm$0.044} & \textbf{11} & ~ & ~\\
        \noalign{\smallskip}
        \citet{2021MNRAS.504..356D} & 31 & \textbf{1719} & \textbf{-3.437$\pm$0.081} & \textbf{-5.406$\pm$0.140} & \textbf{6} & ~ & \textbf{Av: 2.59} \\
        \noalign{\smallskip}
        \citet{2023MNRAS.525.2315A} & ~ & \textbf{1637} &  &  & \textbf{7.4} & ~ & \textbf{Av: 2.689} \\
        \noalign{\smallskip}
        \mbox{\citet{2024A&A...686A..42H}} & 36 & \textbf{1683} & \textbf{-3.453$\pm$0.014} & \textbf{-5.469$\pm$0.012} & \textbf{3.8} & \textbf{-42.48$\pm$21.91} & \textbf{Av: 2.55} \\
        \noalign{\smallskip}\hline
    \end{tabular}
    \begin{tablenotes}
        \item{*}{\bf Column of $E(B-V)$:}
        \item{~} Values in parentheses indicate the color-excess interval when given in the cited papers.
        \item{~} Prefix ``Av:'' to a value means that total extinction in $V$ band is given.
    \end{tablenotes}
\end{table}

\twocolumn[{
    \begin{center}
    \captionof{table}{Literature data on Berkeley 87. Values are formatted as in Table\,\ref{table:A1}.}
    \label{table:A2}
    \footnotesize
    \begin{tabular}{lccccccl}
        \hline\hline  \noalign{\smallskip}

        {Bib source} &  {Number} &
        {distance} &  {pmRA} &  {pmDEC} &  {Age} &  {RadVel} &
        {E(B-V)} \\
    (sorted by year) & {of stars} & (pc) & (mas/yr) & (mas/yr) & (Myr) & (km/s) & (mag) \\
    \hline \noalign{\smallskip} \noalign{\smallskip}
    \citet{1982PASP...94..789T} & 105 & \textbf{946} & ~ & ~ & \textbf{2} & ~ & \textbf{(1.3-1.9)} \\ \noalign{\smallskip}
    \mbox{\citet{1993A&AS..101...37C}} & ~ & 840 & ~ & ~ & 10 & ~ & ~ \\ \noalign{\smallskip}
    \citet{1994RMxAA..29..141F} & ~ & 950 & ~ & ~ & $\leq$10 & ~ & 1.35 \\ \noalign{\smallskip}
    \citet{2000MmSAI..71..687B} & 1 & ~ & ~ & ~ & \textbf{4} & ~ & ~ \\ \noalign{\smallskip}
    \mbox{\citet{2002A&A...390..945K}} & 24 & \textbf{1910} & ~ & ~ & \textbf{3--6} & ~ & \textbf{1.63} \\ \noalign{\smallskip}
    \mbox{\citet{2001A&A...376..441D}} & ~ & \textbf{836} & ~ & ~ & \textbf{10} & ~ & ~ \\ \noalign{\smallskip}
    \citet{2001AJ....121.1050M} & 14 & \textbf{1580} & ~ & ~ & \textbf{3.2} & ~ & \textbf{1.62 (1.4-1.9)} \\ \noalign{\smallskip}
    \mbox{\citet{2002A&A...392..869L}} & 26 & 1910 & ~ & ~ & ~ & ~ & \textbf{(1.5-1.8)} \\ \noalign{\smallskip}
    \citet{2003ARep...47....6L} & 20 & 633 & \textbf{-1.6$\pm$0.67} & \textbf{-4.23$\pm$0.47} & ~ & ~ & ~ \\ \noalign{\smallskip}
    \mbox{\citet{2006A&A...446..949D}} & 47 & ~ & \textbf{-2.53$\pm$0.43} & \textbf{-3.2$\pm$0.43} & ~ & ~ & ~ \\ \noalign{\smallskip}
    \citet{2007BASI...35..383B} & 94 & \textbf{1445} & ~ & ~ & \textbf{1-2} & ~ & ~ \\ \noalign{\smallskip}
    \mbox{\citet{2013A&A...558A..53K}} & ~ & 1239 & \textbf{-5.2$\pm$0.26} & \textbf{-2.4$\pm$0.26} & ~ & -8.6$\pm$1.88 & ~ \\ \noalign{\smallskip}
    \mbox{\citet{2014A&A...564A..79D}} & 99 & ~ & \textbf{-1.93} & \textbf{-3.11} & ~ & -65$\pm$84.8 & ~ \\ \noalign{\smallskip}
    \mbox{\citet{2016A&A...585A.101K}} & ~ & 1239 & ~ & ~ & \textbf{12.6} & ~ & \textbf{1.353} \\ \noalign{\smallskip}
    \citet{2017AstBu..72..257L} & ~ & \textbf{743} & \textbf{-0.678$\pm$0.247} & \textbf{-0.156$\pm$0.303} & \textbf{15} & \textbf{-8.6$\pm$1.88} & \textbf{1.448} \\ \noalign{\smallskip}
    \mbox{\citet{2018A&A...618A..93C}} & 131 & \textbf{1661} & \textbf{-3.654$\pm$0.111} & \textbf{-6.264$\pm$0.144} & ~ & ~ & ~ \\ \noalign{\smallskip}
    \mbox{\citet{2018A&A...619A.155S}} & 1 & 1661 & ~ & ~ & ~ & \textbf{-7.5$\pm$0.37} & ~ \\ \noalign{\smallskip}
    \citet{2018ApJ...864..136B} & ~ & \textbf{1740} & ~ & ~ & ~ & ~ & ~ \\ \noalign{\smallskip}
    \citet{2018MNRAS.478.5184D} & 112 & ~ & \textbf{-3.57} & \textbf{-5.7} & ~ & ~ & ~ \\ \noalign{\smallskip}
    \mbox{\citet{2019A&A...623A..80C}} & 7 & ~ & ~ & ~ & ~ & \textbf{1.98$\pm$2.05} & ~ \\ \noalign{\smallskip}
    \citet{2019ApJS..245...32L} & 95 & ~ & \textbf{-3.664} & \textbf{-6.289} & \textbf{5} & ~ & ~ \\ \noalign{\smallskip}
    \citet{2020MNRAS.495.1209R} & 18 & \textbf{1720} & ~ & ~ & ~ & ~ & ~ \\  \noalign{\smallskip}
    \mbox{\citet{2020A&A...640A...1C}} & 112 & \textbf{1644} & \textbf{-3.654$\pm$0.111} & \textbf{-6.264$\pm$0.144} & \textbf{8.3} & ~ & ~ \\ \noalign{\smallskip}
    \mbox{\citet{2021A&A...647A..19T}} & 1 & \textbf{1590} & ~ & ~ & \textbf{8.5} & \textbf{-7.5$\pm$0.37} & ~ \\ \noalign{\smallskip}
    \mbox{\citet{2021A&A...650A.156D}} & 13 & \textbf{1673} & ~ & ~ & ~ & ~ & 4.7 \\  \noalign{\smallskip}
    \mbox{\citet{2021A&A...651A.104P}} & 110 & \textbf{1644} & \textbf{-3.630$\pm$0.067} & \textbf{-6.321$\pm$0.118} & 8.3 & ~ & ~\\ \noalign{\smallskip}
    \citet{2021ApJ...923..129J} & 148 & \textbf{1669} & \textbf{-3.654$\pm$0.066} & \textbf{-6.268$\pm$0.069} & ~ & ~ & ~ \\ \noalign{\smallskip}
    \citet{2021MNRAS.504..356D} & 130 & \textbf{1708} & \textbf{-3.647$\pm$0.102} & \textbf{-6.250$\pm$0.177} & \textbf{6} & ~ & \textbf{Av: 3.768} \\ \noalign{\smallskip}
    \citet{2023MNRAS.525.2315A} & ~ & \textbf{1665} &  &  & \textbf{8.2} & ~ & \textbf{Av: 3.971} \\ \noalign{\smallskip}
    \mbox{\citet{2024A&A...686A..42H}} & 82 & \textbf{1686} & \textbf{-3.626$\pm$0.008} & \textbf{-6.352$\pm$0.010} & \textbf{5} & \textbf{8.95$\pm$22.08} & \textbf{Av: 4.65} \\
    \noalign{\smallskip}

    \hline
\end{tabular}
\end{center}
}]

\section{Estimation of stellar ages and cluster masses} \label{sec:appB}

To derive stellar ages and masses by interpolating between the two side isochrones, we can use either the
intrinsic CMD with the theoretical tracks transformed to the observational plane or, in reverse, the theoretical
HRD with observational coordinates transformed to $\log L$ (or $M_\mathrm{bol}$) and $\log T_\mathrm{e}$. For the
cluster stars, the values of luminosity were derived as
\begin{equation}
\begin{split}
    \log{L\over{L_\sun}} & =0.4(M_{\rm bol,\,\sun}-M_{\rm bol,*}) \\
    & =0.4(4.73-V_0+DM-BC),
\end{split}
\end{equation}
where $M_{\rm bol,\,\sun}=4.73$ is the absolute magnitude of the Sun (the value recommended by
\citet{2010AJ....140.1158T}), $V_0=V-A_V$ is the intrinsic magnitude of the star taken from photometric catalogs
in Tables \ref{table:3} and \ref{table:4}, $DM$ is the cluster's true distance modulus obtained in the present
paper (Sect.\,\ref{subsec:memb}), and $BC$ is the bolometric correction of the star. To obtain $\log T_\mathrm{e}$
and $BC$, we used the polynomial relations ($\log T_{\rm e}, B-V$) and ($\log T_{\rm e}, BC$) by
\citet{1996ApJ...469..355F}. The colors $B-V$ were taken from tabulations by \citet{Schmidt_Kaler1982} according
to spectral types determined in the {\it Vilnius} photometric system (Tables \ref{table:3} and \ref{table:4}).

The values of $A_V$, $\log T_{\rm e}$, and $\log L$ were also determined using the MK types when available. The
member stars in Berkeley 86 and Berkeley 87 with the parameters derived using the published MK types are listed in
Table\,\ref{MKtable}.

\begin{figure}
   \centering
   \includegraphics[width=\hsize,trim={2.5cm 0.8cm 2.2cm 0.6cm},clip]{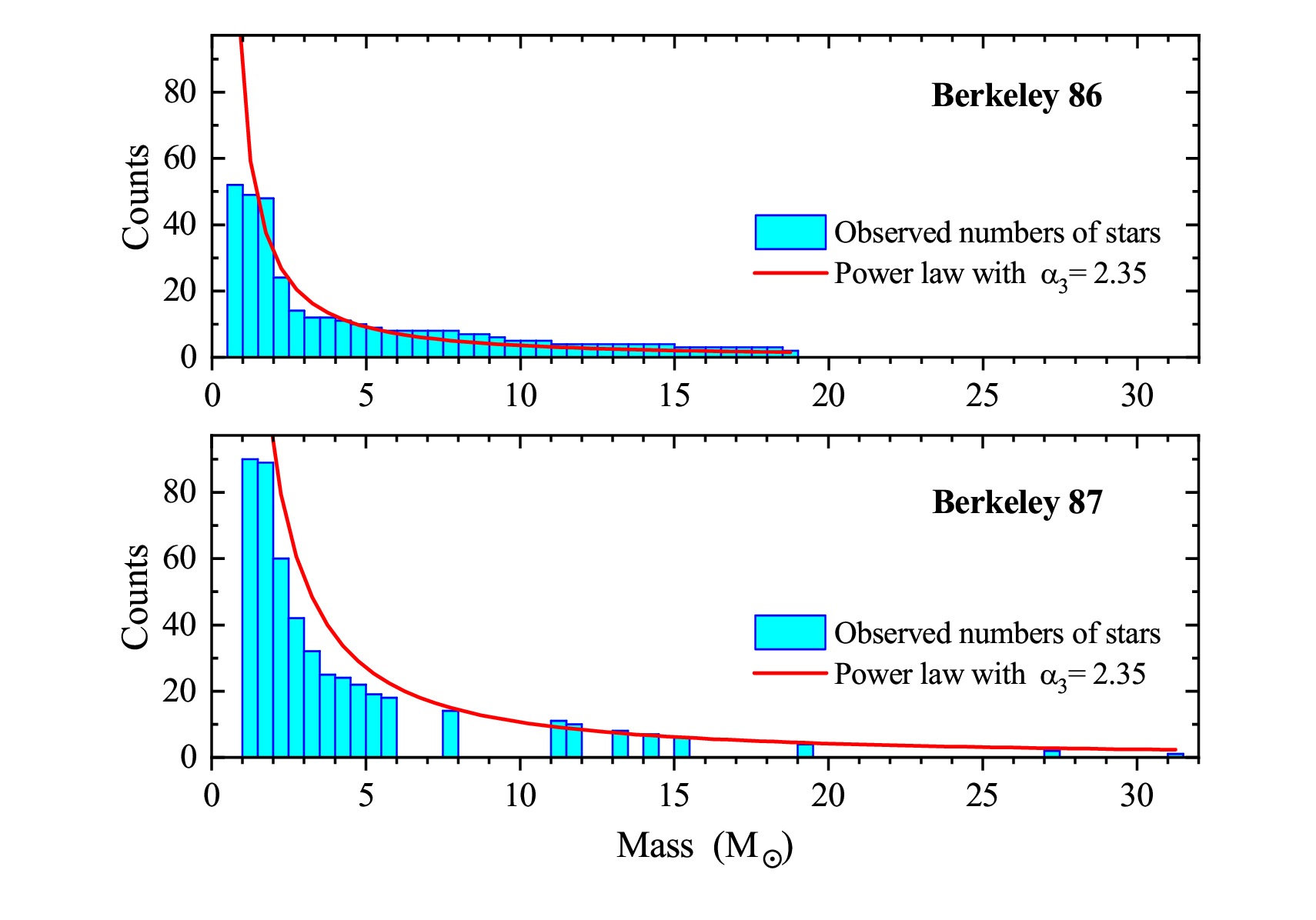}
      \caption{Histogram of the cumulative numbers of observed stars by stellar mass in Berkeley 86 and 87. The line is a
      power law normalized by the total mass of observed stars.}
      \label{fig:MassHist}
   \end{figure}

\begin{figure}
\centering
\includegraphics[width=\hsize,trim={0.4cm 1.6cm 2.0cm 2.5cm},clip]{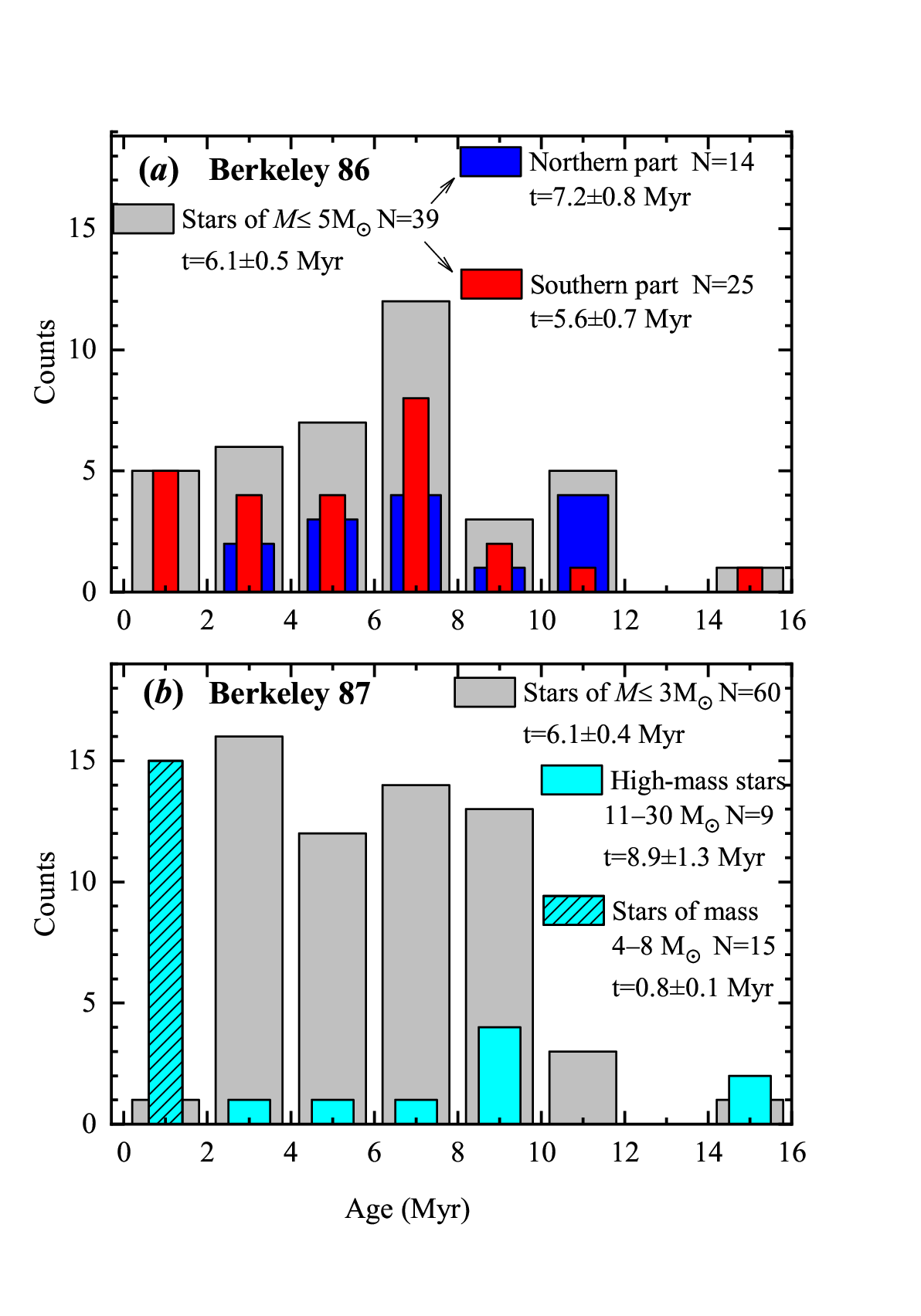}
\caption{The distribution of ages. For Berkeley 86 (panel {\it a}), the bars
    overplotted in blue and red represent stars located in the northern and southern parts of the cluster respectively (see Fig.\,\ref{fig:mass} in Sect.\,\ref{subsec:CMD}); the bars in cyan color are for intermediate-to-high mass stars (they all are located in the southern part).
    For Berkeley 87 (panel {\it b}), the bars in cyan represent separately the high-mass stars (filled bars) and intermediate-mass stars (hatched bar);
    the latter stars all fall in the $<$1 Myr bin. Given in the legends are the median values of ages
    with the $\pm$1$\sigma$ errors evaluted from 16th and 84th percentiles.
} \label{fig:hist_age1}
\end{figure}

\begin{table*}
\centering
\begin{threeparttable}
    \caption[]{Member stars with MK spectral types from the literature.} \label{MKtable}
    \centering                         
    \begin{tabular}{r l l l l l l l r r r r}
        \hline\hline \noalign{\smallskip}
        \multicolumn{1}{l}{Star} & \multicolumn{1}{c}{Simbad ID} & \multicolumn{3}{c}{MK spectral type} &
        \multicolumn{1}{c}{$(B-V)_0$} & \multicolumn{1}{c}{$\log T_{\rm e}$} & \multicolumn{1}{c}{$\log L$} &
        \multicolumn{2}{c}{Mass$^\dag$} & \multicolumn{1}{c}{Age}\\
        & & \multicolumn{1}{l}{Massey et al.$^{(1),(2)}$} & \multicolumn{2}{l}{Other references} & & & &
        \multicolumn{2}{c}{(${\rm M}_\sun$)} & \multicolumn{1}{c}{(Myr)}\\
        \hline \noalign{\smallskip} \noalign{\smallskip}
        \multicolumn{1}{c}{\rlap{\it{Berkeley~86}}} & & & & & & & & & &\\
        \noalign{\smallskip}
        287 & LS\,II~+38~49 & B0.2\,IV & & & --0.29 & 4.488 & 4.332 & 15.01 & {\it 13.38} &\\
        353\rlap{*} & EM*VES 201A & & B1\,Ve & (6) & --0.26 & 4.386 & 3.653 & 9.14 & {\it 7.61} &\\
        476\rlap{*} & Cl Berkeley 86~3 & & B1\,V & (4) & --0.26 & 4.386 & 4.134 & 10.92 & {\it 11.78} & 15.41\\
        528\rlap{*} & HD~228989 & O9.5\,V\,+\,O9.5\,V & O8.5\,V\,+\,O9.7\,V & (7) & --0.30 & 4.527 & 4.639 & 18.89 & {\it
            17.11} &\\
        898 & [NH52]~38 & B1.5\,V & & & --0.25 & 4.356 & 3.618& 8.39 & {\it 7.81} & 13.43\\
        \noalign{\smallskip} \noalign{\smallskip}
        \multicolumn{1}{c}{\rlap{\it{Berkeley~87}}} & & & & & & & & &\\
        \noalign{\smallskip}
        24 & HD~229059 & B1\,Ia & B2\,Iabe & (3) & --0.19 & 4.309 & 5.528 & 30.63 & & 5.35\\
        54 & LS~II +37~73 & B0.2\,III & B0.5\,Ve & (6)& --0.29 & 4.488 & 4.857 & 19.35 & & 6.65\\
        106 & EM*VES~203 & B0.5\,V & B0.5\,Ve &(6) & --0.28 & 4.452 & 4.348 & 13.65 & {\it 12.81} & 8.13\\
        139 & EM*AS 407 & & B2\,e & (6) & --0.24 & 4.328 & 3.689 & 8.61 & {\it 7.25} &\\
        168 & Cl Berkeley 87~13 & B0.5\,III: & & & --0.28 & 4.452 & 4.407 & 14.02 & & 8.59\\
        179 & Cl Berkeley 87~16 & B2\,V & & & --0.24 & 4.328 & 3.483 & 7.60 & {\it 7.16} &\\
        234 & BD +36 4032 & O8.5\,III & O8.5\,V,~S\rlap{B} & (5) & --0.31 & 4.567 & 5.290 & 30.66 & & 3.36\\
        243\rlap{*} & Cl Berkeley 87~26 & B0.5\,I & & & --0.28 & 4.452 & 4.557 & 15.19 & & 9.00\\
        285\rlap{*} & WR~142 & & WO2 & (8) & & 5.30 & 5.35 & 11.88 & & \\
        &        & & & & & 5.22 & 3.91 & 2.53 & &\\
            298 & Cl Berkeley 87~31 & B1\,V & & & --0.26 & 4.386 & 4.161 & 11.06 & {\it 11.50} & 15.40\\
        314 & Cl Berkeley 87~32 & B0.5\,III & & & --0.28 & 4.452 & 4.587 & 15.32 & & 9.00\\
            413 & EM*VES~204 & B2\,III: & B1\,Ve & (6) & --0.26 & 4.386 & 4.264 & 11.70 & {\it 12.41} & 15.00\\
        BC Cyg\rlap{*} & BD +37 3903 & & M3.5\,Ia & (9) &  & 3.548 & 5.266 & 19.74 & & 8.36\\
            \noalign{\smallskip} \hline
        \end{tabular}
        \begin{tablenotes}
            {\footnotesize
                \item{~}{\bf References for MK types:} (1) \citet{1995ApJ...454..151M} for Be\,86; (2) \citet{2001AJ....121.1050M} for Be\,87;
                (3) \citet{1982PASP...94..789T}; (4) \citet{1992AJ....103..916F}; (5) \citet{2004AN....325..380N};
                (6) \citet{2011BASI...39..517M}; (7) \citet{2013A&A...550A..27M}; (8) \citet{1995A&A...295...75K}; \citet{1977ApJ...218..181F}.
                \item{*}{\bf Notes on stars:}
                \item{~}Be\,86-476. SB2 \citep{2006ApJ...648..580H}.
                \item{~}Be\,86-528. The parameters are given for one of the two components of
                equal brightness, according to MK type from Ref. (1).
                \item{~}Be\,87-243. Luminosity class given in Ref. (2) is incompatible with that from {\it
                    Gaia} DR3 parallax, therefore the parameters are derived using luminosity class III confirmed by photometric {\it Vilnius} classification.
                \item{~}Be\,87-285 (WR 142). The values of $\log T_{\rm e}$ and $\log L$ for this WR star are from
                \citet{2015A&A...581A.110T} ({\it first line}) and from \citet{2024MNRAS.531.2422T}, as cited in their Table 1
                ({\it second line}). Both values of luminosity are adjusted to match our distance modulus for Berkeley~87
                ($DM=11.13$ mag); accordingly, the mass is updated by using the $M/L$ relation from \citet[][Eq.19]{1989A&A...210...93L}.
                \item{~}BC Cyg. Cluster member \citep{2024A&A...686A..42H}. $\log T_{\rm e}$ and $\log L$ are from
                \citet{2020A&A...644A..62C}, the latter value is adjusted to $DM=11.13$ mag.
                \item{$\dag$} {\bf Masses} in the left hand column are determined by interpolating data between the Padova $Z=0.0152$ theoretical
                isochrones \citep{2002A&A...391..195G, 2012MNRAS.427..127B}. Masses given in italics
                (right-hand column) are for the main-sequence stars, determined using the classical mass-luminosity relation from \citet{2018MNRAS.479.5491E}.
            }
        \end{tablenotes}
    \end{threeparttable}
\end{table*}

For the calculation of cluster masses we use a multiple power-law IMF \citep{2004MNRAS.348..187W}:
\begin{equation}
    \xi(m) = k\left\{
    \begin{array}{ll}
        \left(\frac{m}{m_\mathrm{H}}\right)^{-\alpha_1}, & \quad m_\mathrm{H}\leq m\leq m_0,\\
        \left(\frac{m_0}{m_\mathrm{H}}\right)^{-\alpha_1} \left(\frac{m}{m_0}\right)^{-\alpha_2}, & \quad m_0\leq m\leq m_1,\\
        \left(\frac{m_0}{m_\mathrm{H}}\right)^{-\alpha_1} \left(\frac{m_1}{m_0}\right)^{-\alpha_2} \left(\frac{m}{m_1}\right)^{-\alpha_3}, & \quad m_1\leq m\leq m_\mathrm{max},\\
    \end{array}
    \right.
\end{equation}
with exponents
\begin{equation}
    \begin{array}{ll}
        \alpha_1 = 1.30 & 0.08\leq m/\mathrm{M}_\sun \leq0.50,\\
        \alpha_2 = 2.30 & 0.50\leq m/\mathrm{M}_\sun \leq1.00,\\
        \alpha_3 = 2.35 & 1.00\leq m/\mathrm{M}_\sun.\\
    \end{array}
\end{equation}
The normalization constant $k$ is given by the mass of the cluster in the stellar mass range between $m_\mathrm{min}$ and $m_\mathrm{max}$
\begin{equation}
    M^\mathrm{cl}=\int_{m_\mathrm{min}}^{m_\mathrm{max}}m\xi(m)dm,
\end{equation}
which can be found by means of an additional equation
\begin{equation}\label{eqn:Mcl}
    {\sum_{m=m_2}^{m=m_\mathrm{max}} m_i}=k \left(\frac{m_\mathrm{H}}{m_0}\right)^{\alpha_1} \left(\frac{m_0}{m_1}\right)^{\alpha_2} m_1^{\alpha_3}\left(\frac{m_\mathrm{max}^{2-\alpha_3}-m_2^{2-\alpha_3}}{2-\alpha_3}\right)
\end{equation}
for $m_\mathrm{max}> m_2>m_1$.

Having the total cluster mass calculated, it is possible to estimate its tidal radius at which the cluster's
density declines steeply and drops to zero. As argued in \citet{2008gady.book.....B}, $r_\mathrm{t}$ can be
approximately identified with the Jacobi radius \citep[see, also Eq.\,(7)][]{1962AJ.....67..471K}:
\begin{equation}\label{eqn:rJ}
    r_\mathrm{J}=\left[\frac{GM}{4A(A-B)}\right]^{1/3},
\end{equation}
where $A$ and $B$ are Oort's constants at the Galactocentric distance of the cluster with the total mass $M$
enclosed within $r_\mathrm{J}$, and $G$ is the gravitational constant. The values of $r_\mathrm{J}$ in given
Table\,\ref{table:B2} are calculated with $(A-B)$\,=29.8 {\kms}\,kpc$^{-1}$ and $A$\,=15.6 {\kms}\,kpc$^{-1}$,
corresponding to a `universal' Galactic rotation curve by \citet{2019ApJ...885..131R}.


\begin{table}
    \begin{threeparttable}
        \caption[]{Normalization of the cluster stellar mass function and the calculations of cluster masses in the
            observed fields of Berkeley 86 and Berkeley 87. $N$ is the number of stars used in summing up stellar masses
            from $m_2$ to $m_\mathrm{max}$. The last two columns list the total cluster mass down to
            0.08\,$\mathrm{M}_\sun$ and Jacobi radius.
        }\label{table:B2}
        \centering                         
\begin{tabular}{lrrrrrr}
            \hline\hline \noalign{\smallskip}
            \multicolumn{1}{c}{Cluster} & \multicolumn{1}{c}{$N$} & \multicolumn{1}{c}{$m_\mathrm{max}$}&
            \multicolumn{1}{c}{$m_2$} & \multicolumn{1}{c}{$\Sigma m_i$} &
            \multicolumn{1}{c}{$M^\mathrm{cl}$} &
            \multicolumn{1}{c}{$r_\mathrm{J}$~(pc)}\\
            \noalign{\smallskip}
            \hline \noalign{\smallskip} \noalign{\smallskip}
            \noalign{\smallskip}
            Berkeley 86 & 44 & 18.9 & 1.73 & 178.7 & 519 & 10.6\\
                        & 11 & 18.2 & 4.38 & 111.3 & 633 & 11.3\\
                        \noalign{\smallskip}
            \hline \noalign{\smallskip} \noalign{\smallskip}
            Berkeley 87$^*$ & 13 & 31.4 & 7.53 & 193.1 & 1408 & 14.8\\
                        & 14 & 27.3 & 7.53 & 211.9 & 1656 & 15.6\\
            \noalign{\smallskip}
            BC Cyg & 14 & 31.4 & 7.53 & 212.8 & 1551 & 15.3\\
            included & 15 & 27.3 & 7.53 & 231.6 & 1811 & 16.1\\
            \noalign{\smallskip} \noalign{\smallskip}
            \noalign{\smallskip} \hline
        \end{tabular}
        \begin{tablenotes}
            \footnotesize
            \item{*} For Berkeley 87, the normalization constant $k$ and the cluster's mass are calculated using only
            brighter end of the observed MF, since the {\it Vilnius} CCD photometry covers the cluster's central part where high-mass stars are
            concentrated (Fig.\,\ref{fig:mass} in Sect.\,\ref{subsec:CMD}). Outside the CCD field such stars are almost absent, except
            for the high-mass member star BC Cyg \citep{2024A&A...686A..42H} of $\sim$\,19 M$_\sun$
            \citep{2020A&A...644A..62C}. First and second rows refer to different status of the massive
            cluster member No.\,234: when it is taken as single and the most massive (first row) or adopted as having equal mass
            components (second row) which then become next to star No.\,24 (27 $\mathrm{M}_\sun$) by mass.
        \end{tablenotes}
    \end{threeparttable}
\end{table}

\section{Radial velocity data}  \label{sec:appC}

The average values of RVs for Be\,86 and Be\,87, given in different literature sources (see Tables \ref{table:A1}
and \ref{table:A2}), show disagreement from one source to another, and, in fact, any of these values can hardly be
used as representative of the clusters' RVs. Therefore, prior to adopting truly satisfactory averages, we had
first to collect and examine the RV data available for the individual member stars.

Most of the RV data collected come from the {\it Gaia} DR3, APOGEE DR17 \citep{2022ApJS..259...35A}, APOGEE DR14
\citep{2019A&A...623A..80C}, and {LAMOST DR5} \citep{2020A&A...640A.127Z} surveys. For the member stars with the
APOGEE DR17 data available, we used the recent RV catalog by \citet{2025AJ....169..167S}, based on a new reduction
of APOGEE DR17 visits, that offers improved estimates of RVs and well-calibrated uncertainties. We have also
collected RVs from earlier literature sources that give data for a number of member stars in Berkeley 86
\citep{1992AJ....103..916F,2006ApJ...648..580H}, NGC\,6913 \citep{2004A&A...415..145B}, and IC\,4996
\citep{1999AJ....118.1759D,2010ApJ...722..605H}. The systemic velocities for known SB stars were taken from
orbital solutions by \citet{2013A&A...550A..27M} and \citet{2004A&A...415..145B}. Many of the stars collected have
multiple observations at different epochs, with a major contribution from APOGEE and {\it Gaia} surveys and, to a
much lesser extent, from other literature sources.

The majority of the stars with RV data are mainly in the spectral range O8.5--B, while only very few are classified
as of types A to F, plus one red supergiant. Although the number of these stars in each of the clusters is not small (from
12 in Berkeley 86 to $\sim$30 in Berkeley 87 and NGC~6913), a significant fraction of them have grossly discrepant RV
values, in some cases exceeding 100 {\kms}, or show the measurement errors larger than 10 {\kms}. Much of the
scatter in the RVs can be accounted for by unrecognized binarity (e.g., \citet{2022ApJS..259...19L}  estimate the fraction of
binaries in most of the OCs to vary from 30\% to 55\%). An additional contribution to the scatter may come from other
sources of variability, such as rapid rotation, stellar pulsations or strong stellar wind, that may occur in early type
stars.

\begin{figure}
    \centering
    \includegraphics[width=\hsize,trim={0.8cm 1.2cm 1.4cm 2.0cm},clip]{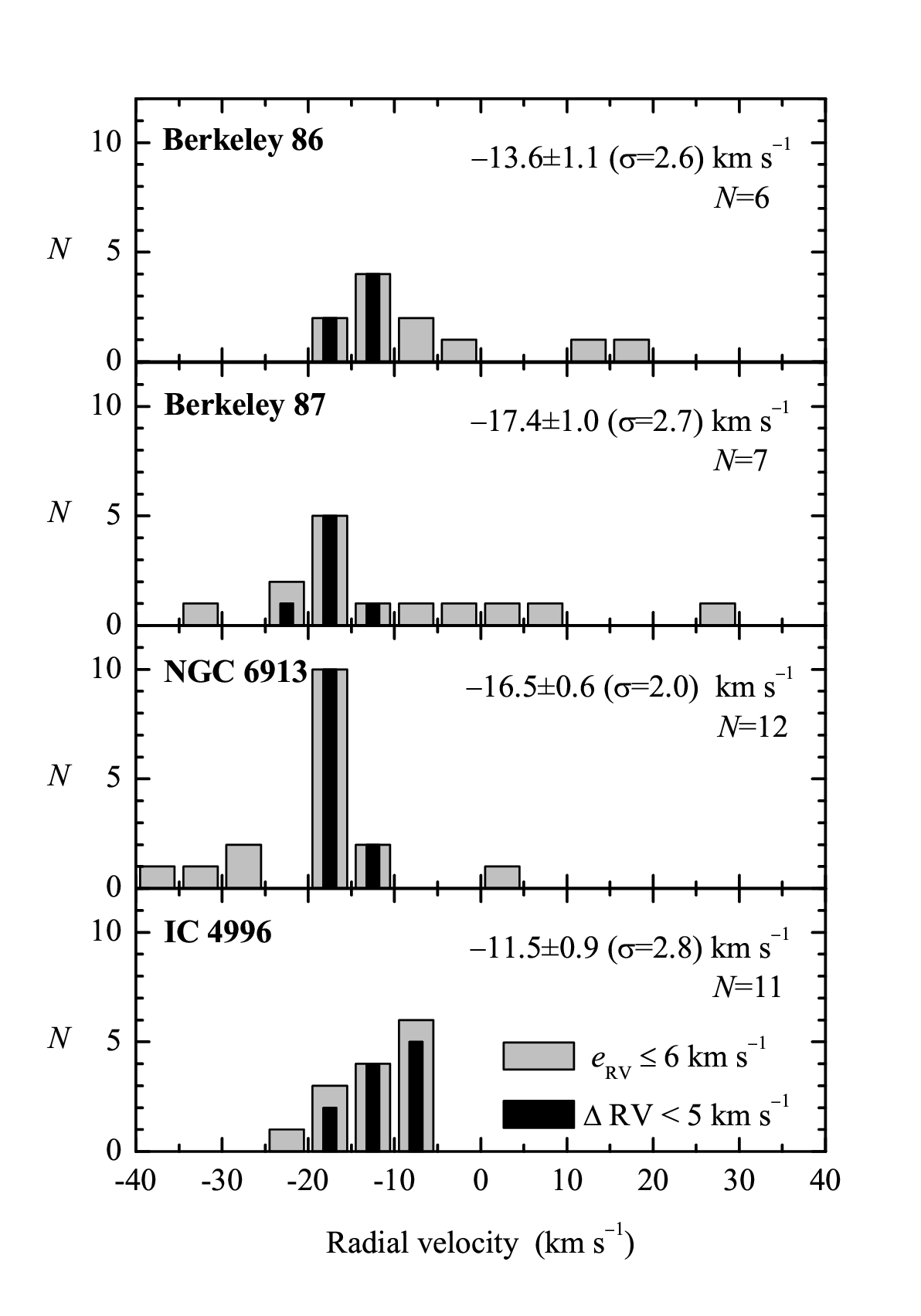}
    \caption{The distributions of radial velocities with measurement errors of $\leq 6$ {\kms} for the member stars
        in Berkeley 86, Berkeley 87, NGC 6913, and IC 4996. Overplotted black bars indicate the spread in
        RVs of the stars used to obtain average RVs (the values are given in the upper-right corner
        of each panel).
    } \label{fig:RV}
\end{figure}

In order to derive a reliable central RV value for each cluster, we only considered those stars for which the
error of a weighted mean (in the case of multiple RV observations) or of single RV measurement was $\leq$6 {\kms}.
Such an upper bound for the RV error was chosen in order not to miss the RV data for a number of member stars in
Berkeley 86 that were provided by \citet{2006ApJ...648..580H} with a typical uncertainty of 6 {\kms}. Although the
removal of stars with inaccurate RVs has led to a much smaller RV spread for the same cluster, the resulting RV
distributions are thinly stretched and asymmetric (see the histograms in Fig.\,\ref{fig:RV}), thus leaving us
without a possibility of averaging RVs directly or fitting the distributions by a Gaussian profile. Therefore, for
each cluster we finally adopted the average calculated by iteratively discarding the most deviating outliers until
the RVs of the stars left deviated from their mean by less than 5 {\kms}. The adopted average values of RVs are
indicated in the legends of Fig.\,\ref{fig:RV}. It should be noted that for Berkeley 87, all RV stars used in
averaging, except one, are from the APOGEE survey.

In Berkeley 86, one spectroscopic binary, No.\,528 (HD 228989),  has its orbital solution.
\citet{2013A&A...550A..27M} give the systemic velocity $-3.7\pm1.8$ {\kms} and $-7.8\pm1.9$ {\kms} for the primary
and the secondary components, respectively, and the mass ratio 1.13. These values differ from the cluster's
average RV adopted above by several {\kms}. \citet{2013A&A...550A..27M} determined spectral types of O\,8.5\,V and
O\,9.7\,V for the components and admitted them to be very close to fill their Roche lobe. A difference of 4 {\kms}
between the values of systemic velocity of the components can indicate that their spectral lines were differently
affected by the velocity field. Therefore we had no firm reason to adopt for Be\,86 any of the two systemic
velocity values of this only one star as the cluster's overall RV, even though the secondary's value, within the
errors, does not drastically differ from the cluster's average RV. In the case of NGC~6913, its average RV value
agrees well with the systemic velocity of the binary member star NGC~6913~9 of B2\,V type, $-17.0\pm6.0$ {\kms}
\citep{2004A&A...415..145B}.


\section{Calculation of Galactocentric velocities} \label{sec:appD}

The peculiar velocities $(U,V,W)$ with respect to the Local Standard of Rest (LSR) are calculated in a
right-handed Cartesian coordinate system $(X,Y,Z)$ centered at the Sun. For calculations of the velocity
components and their errors we used the matrix equations given by \citet{1987AJ.....93..864J}, modified to accept
positions for equinox 2000.0, and the values of the Sun's velocity with respect to LSR, $(U_\sun,V_\sun,W_\sun)=
(11.1,12.2,7.2)$ {\kms}, from \citet{2010MNRAS.403.1829S}.

Space velocities $(U,V,W)$ with respect to the Local Standard of Rest (LSR) can be transformed to peculiar
velocities $(U_\mathrm{s},V_\mathrm{s},W_\mathrm{s})$ with respect to regional standard of rest (RSR) by rotating
the vector $(U,V,W)$ through the Galactocentric azimuth angle of a given source and removing circular velocity of
Galactic rotation at the Galactocentric distance of that source, following the formulae given in
\citet{2009ApJ...700..137R}:
\begin{eqnarray}
U_\mathrm{s}&=&U\cos\beta -(V+\theta_0)\sin\beta\,,\\
V_\mathrm{s}&=&U\sin\beta+(V+\theta_0)\cos\beta-\theta\,,\\
W_\mathrm{s}&=&W\,.
\end{eqnarray}
Here, the angle $\beta$ is the Galactocentric azimuth, the component $U_\mathrm{s}$ is directed radially toward the Galactic
center, the azimuthal component $V_\mathrm{s}$ is positive in the local direction of
Galactic rotation, and the vertical component $W_\mathrm{s}$, directed toward the north Galactic pole, is the same
as $W$. For transformation we adopted a `universal' Galactic rotation curve by \citet{2019ApJ...885..131R}, with
the Sun's Galactocentric distance  $R_0=8.15\pm0.15$ kpc and the rotation velocity at the solar circle
$\theta_0=236\pm 5$ {\kms}. At the Galactocentric distances of the open clusters considered here, this rotation
curve gives a circular velocity $\theta$ by only $0.3-0.4$ {\kms} higher than $\theta_0$ (all modern rotation
curves are nearly flat within a few kiloparsecs of the Sun).

The values of $U_\mathrm{s}$ and $V_\mathrm{s}$ are dependent on the adopted parameters $R_0$, $\theta_0$ and
$V_\sun$ which are not precisely known. Any change in the latter two parameters would lead to a systematic change
in $V_\mathrm{s}$. The component $U_\mathrm{s}$, however, is much less susceptible to uncertainties of $\theta_0$
and $V_\sun$. In addition, uncertainties in the adopted values of heliocentric RVs can also be important sources
of systematic error in the azimuthal velocity $V_\mathrm{s}$. The clusters given in
Table\,\ref{table:orbit_params} (the key parameters are also summarized in Table\,\ref{uvw}) are located nearly at
tangent point (their galactic latitude, $\ell$$\simeq$76\degr, and the Galactocentric azimuth, $\simeq$12\degr,
constitute the angle close to $90\degr$), hence their heliocentric RVs measure almost entirely the azimuthal
motion (including the systematic effect of Galactic rotation), whereas the other two velocity components,
$U_\mathrm{s}$ and $W_\mathrm{s}$, remain almost independent on RV. Using the Galactic rotation curve by
\citet{2019ApJ...885..131R} the systematic effect in the observed RVs due to Galactic rotation is expected to be
of the order of $\sim$7 {\kms}.


\begin{table*}
\begin{threeparttable}
    \centering
    \caption{Input parameters, space velocities, and orbital properties of four open clusters in Cygnus.}
    \label{table:orbit_params}
    \begin{tabular}{lccccc}
        \hline \hline
        \noalign{\smallskip}
        \textbf{Parameter} &  \multicolumn{4}{c}{\textbf{Clusters}}  & \textbf{Units} \\
        \hline
        \noalign{\smallskip}
        Simbad Name & Berkeley 86 & Berkeley 87 & NGC 6913 & IC 4996 & kpc \\
        \noalign{\smallskip}
        \textit{Cluster parameters\textsuperscript{a}} &  &  &  & & \\
        Age ($T$) & 6 & 7 & 8.3$^\ast$ & 10.6$^\ast$ & Myr \\
        Number member stars ($N$) & 63 & 92 & 252 & 212 & \dots \\
        \noalign{\smallskip}
        \textit{Heliocentric inputs\textsuperscript{a}} &  &  &  & & \\
        Right ascension ($\alpha_{2000}$) & $305.074 \pm0.007$ & $305.410 \pm0.008$ & $305.898 \pm0.013$  & $304.145 \pm0.006$ & $\deg$ \\
        Declination ($\delta_{2000}$) & $+38.714 \pm0.010$ & $+37.402 \pm0.006$ & $+38.521 \pm0.012$  & $+37.634 \pm0.005$ & $\deg$ \\
        Distance ($d$) & $1691 \pm 9$  &  $1681 \pm 9$ & $1694 \pm 7$ & $1954 \pm 5$$^\dag$ & pc \\
        Radial velocity ($v_{r}$)\textsuperscript{b} & $-13.6 \pm 1.1$ (6)  & $-17.4 \pm 1.0$ (7) & $-16.5 \pm 0.6$ (12)  & $-11.5 \pm 0.9$ (11) & km s$^{-1}$ \\
        Proper motion ($\mu_{\alpha} \cos \delta$) & $-3.487 \pm 0.018$  & $-3.631 \pm 0.010$ & $-3.339 \pm 0.016$  & $-2.631 \pm 0.006$$^\dag$ & mas yr$^{-1}$ \\
        Proper motion ($\mu_{\delta}$) & $-5.454 \pm 0.017$  & $-6.351 \pm 0.013$ & $-5.794 \pm 0.013$  & $-5.365 \pm 0.007$$^\dag$ & mas yr$^{-1}$ \\
        \noalign{\smallskip}
        \textit{Tangential velocities} &  &  &  & & \\
        Tangential velocity ($v_\ell$) & $-51.9 \pm 0.3$ & $-58.1 \pm 0.3$ & $-53.6 \pm 0.3$ & $-54.7 \pm 0.2$ & km s$^{-1}$ \\
        Tangential velocity ($v_b$) & $-1.7 \pm 0.1$ & $-5.0 \pm 0.1$ & $-4.6 \pm 0.1$ & $-7.4 \pm 0.1$ & km s$^{-1}$ \\
        \noalign{\smallskip}
        \textit{Velocities with respect to LSR\textsuperscript{c}}  &  &  & & & \\
        Velocity ($U$) & $58.4 \pm 0.8$ & $63.1 \pm 0.9$ & $59.6 \pm 0.8$ & $61.2 \pm 0.8$ & km s$^{-1}$ \\
        Velocity ($V$) & $-12.9 \pm 1.1$  & $-18.9 \pm 1.1$ & $-16.0 \pm 0.7$ & $-12.6 \pm 1.0$ & km s$^{-1}$ \\
        Velocity ($W$) & $5.2 \pm 0.5$ & $2.1 \pm 0.5$ & $2.5 \pm 0.4$ & $-0.4 \pm 0.4$ & km s$^{-1}$ \\
        \noalign{\smallskip}
        \textit{Velocities with respect to RSR\textsuperscript{d}}  &  &  & & & \\
        Velocity ($U_\mathrm{s}$) & $10.9 \pm 0.4$ & $17.0 \pm 0.4$ & $12.5 \pm 0.5$ & $5.9 \pm 0.4$ & km s$^{-1}$ \\
        Velocity ($V_\mathrm{s}$) & $-6.0 \pm 1.4$ & $-10.9 \pm 1.4$ & $-8.7 \pm 1.0$ & $-4.8 \pm 1.2$ & km s$^{-1}$ \\
        Velocity ($W_\mathrm{s}$) & $5.2 \pm 0.5$ & $2.1 \pm 0.5$ & $2.5 \pm 0.4$ & $-0.4 \pm 0.4$ & km s$^{-1}$ \\
        \noalign{\smallskip}
        \textit{Galactocentric parameters\textsuperscript{d}}  &  &  & & & \\
        Current radius ($R$) & $7.93 \pm 0.15$ & $7.91 \pm 0.15$ & $7.94 \pm 0.14$ & $7.89 \pm 0.14$ & kpc \\
        Perigalaxy ($R_{\mathrm{peri}}$) & $7.36\pm0.15$ & $6.97\pm0.13$ & $7.19\pm0.14$  & $7.46\pm0.15$ & kpc \\
        Apogalaxy ($R_{\mathrm{apo}}$) & $8.06\pm0.18$ & $8.10\pm0.18$ & $8.08\pm0.18$  & $7.93\pm0.18$ & kpc \\
        Eccentricity ($e$) & $0.045 \pm 0.006$ & $0.075 \pm0.007$ & $0.058 \pm0.005$  & $0.030\pm0.006$ & \dots \\
        Coordinate ($X(t_{\mathrm{birth}})$)  &$7.99\pm0.16$ & $8.01\pm0.16$ & $8.02\pm0.16$ & $7.91\pm0.17$ & kpc \\
        Coordinate ($Y(t_{\mathrm{birth}})$)   &$0.255\pm0.047$ & $0.049\pm0.055$ & $-0.253\pm0.060$ & $-0.591\pm0.087$ & kpc \\
        Coordinate ($Z(t_{\mathrm{birth}})$)  &$21\pm2$ & $11\pm2$ & $12\pm2$ & $47\pm2$ & pc \\
        Coordinate ($X(t_{\mathrm{age}})$)  &$7.77\pm0.16$ & $7.74\pm0.15$ & $7.77\pm0.15$ & $7.76\pm0.16$ & kpc \\
        Coordinate ($Y(t_{\mathrm{age}})$)   &$1.645\pm0.009$ & $1.629\pm0.009$ & $1.650\pm0.007$ & $1.890\pm0.005$ & kpc \\
        Coordinate ($Z(t_{\mathrm{age}})$)  &$58\pm2$ & $30\pm2$ & $39\pm2$ & $64\pm2$ & pc \\
        Max. vertical height ($Z_{\mathrm{max}}$)  & $84.7\pm2.3$ & $37.4\pm1.7$ & $47.7\pm2.0$ & $63.2\pm1.9$ & pc \\
        \noalign{\smallskip}
        \textit{Dynamical timescales}\textsuperscript{d}  &  &  & & & \\
        Time passing Galaxy plane ($t_{Z=0}$)  &  $-9.2\pm0.3$ & $-10.6\pm0.6$ & $-11.6\pm0.5$  & $-20.5\pm0.8$ & Myr \\
        Radial period ($P_R$) &  $148.4\pm5.3$ & $144.3\pm5.2$ & $146.7\pm4.9$  & $147.6\pm5.9$ & Myr \\
        Vertical period ($P_z$) &  $71.4\pm2.3$ & $67.2\pm2.0$ & $69.4\pm2.0$  & $73.8\pm3.4$ & Myr \\
        \hline
    \end{tabular}
    \begin{tablenotes}
        \footnotesize
        \item{$^\ast$} The ages for NGC 6913 and IC\,4996 are from \citet{2023MNRAS.525.2315A}.
        \item{$^\dag$} For IC\,4996, the input parameters $r$ and $\mu$ are from the catalog of \citet{2024A&A...686A..42H}.
        \item{\textsuperscript{a}} Derived from refined parameters (this work).
        \item{\textsuperscript{b}} In parentheses are the number of stars used for RV averaging.
        \item{\textsuperscript{c}} Calculated with the Sun's velocity $(U_\sun,V_\sun,W_\sun)=(11.1,12.2,7.2)$ {\kms} from Sch{\"o}nrich et al.
        (2010).
        \item{\textsuperscript{d}} Calculated using the Galactic rotation curve from \citet{2019ApJ...885..131R} with
        $R_0=8.15\pm 0.15$ kpc and $\theta_0=236\pm 7$ {\kms}, and adopting $z_{\odot} = 20.8$ pc \citep{2016ARA&A..54..529B}.
        The errors given do not include the contribution from uncertainties in the adopted parameters of the Galaxy and
        model potentials.
    \end{tablenotes}
\end{threeparttable}
\end{table*}

\end{appendix}

\end{document}